\documentclass[10pt,twocolumn,twoside,journal]{IEEEtran}
\usepackage{amsmath}
\usepackage{times}
\usepackage{amsbsy,color}
\usepackage{latexsym}
\usepackage{amssymb}
\usepackage{graphicx}
\usepackage{subfigure}
\usepackage{amsfonts}
\usepackage{epsfig,subfigure,tikz}
\allowdisplaybreaks[4] %
\begin{document}

\title{\huge Distributed Low-Rank Adaptive Algorithms Based on Alternating Optimization and Applications}
\author{Rodrigo C. de~Lamare  
\thanks{R. C. de Lamare is with CETUC / PUC-Rio, Brazil and the Department of Electronics, University of York, U.K.
(e-mail: rodrigo.delamare@york.ac.uk)}.} \maketitle

\begin{abstract}
This paper presents a novel distributed low-rank scheme and adaptive
algorithms for distributed estimation over wireless networks. The
proposed distributed scheme is based on a transformation that
performs dimensionality reduction at each agent of the network
followed by transmission of a reduced set of parameters to other
agents and reduced-dimension parameter estimation. Distributed
low-rank joint iterative estimation algorithms based on alternating
optimization strategies are developed, which can achieve
significantly reduced communication overhead and improved
performance when compared with existing techniques. A computational
complexity analysis of the proposed and existing low-rank algorithms
is presented along with an analysis of the convergence of the
proposed techniques. Simulations illustrate the performance of the
proposed strategies in applications of wireless sensor networks and
smart grids.

\end{abstract}



\section{Introduction}

\IEEEPARstart{D}{istributed} strategies have become fundamental for
parameter estimation in wireless networks and applications such as
sensor networks \cite{Lopes1,Lopes2,xu2015adaptive} and smart grids
\cite{Xie,Xu2}. Distributed techniques deal with the extraction of
information from data collected at nodes that are distributed over a
geographic area \cite{Lopes1}. In this context, a specific sensor
node or agent in the network collects processed data from its
neighbors and combines them with its local information to generate
improved estimates. However, when the unknown parameter vector to be
estimated has a large number of parameters, the network requires a
large communication bandwidth between neighboring nodes to transmit
their local estimates. This problem limits the usefulness of
existing algorithms in applications with large data sets as the
convergence speed is dependent on the number of parameters
\cite{Lopes2,Haykin,Sayed}. Hence, distributed dimensionality
reduction has become an important tool for distributed inference
problems with large data sets.

In order to perform dimensionality reduction or compression, several
algorithms have been proposed in the literature in the context of
distributed quantized Kalman filtering \cite{Msechu,Xiao1},
quantized consensus algorithms \cite{Pereira}, distributed principal
subspace estimation \cite{Li3}, the single bit strategy
\cite{Sayin1} and Krylov subspace optimization techniques
\cite{Chouvardas1}. However, these distributed algorithms
\cite{Msechu}-\cite{Chouvardas1} have drawbacks such as high
computational complexity, unsatisfactory performance and
implementation issues. Available distributed approaches for
dimensionality reduction or compression
\cite{Msechu}-\cite{Chouvardas1} have trade-offs between the amount
of cooperation, communication and system performance. This calls for
the development of cost-effective techniques that can approach the
performance of theoretical bounds for parameter estimation, have
flexibility and high-compression capability, and exhibit low
computational complexity. In this context, low-rank techniques are
powerful tools to perform dimensionality reduction, which have been
applied to spread-spectrum systems
\cite{Scharf_bv,Scharf,Honig02,aifir,Lamare,echojidf,jidf,jiols,sjidf,doajio,wljio,wlmwf,Clarke,barc,locsme,doajidf,okspme,wljidf},
multi-input-multi-output (MIMO) systems
\cite{mmimo,Sun,wence,Lamare1,gbd,mbthp,rmbthp,mberjidf,wlrbd} and
beamforming \cite{Qian03,Nuan,Wang1}. However, limited research has
been carried out on distributed low-rank estimation, in which the
distributed principal subspace estimation \cite{Li3} and the Krylov
subspace optimization \cite{Chouvardas1} techniques are recent
contributions. {Related approaches to low--rank techniques include
compressive sensing-based strategies \cite{Yao},\cite{dce}, which
exploit sparsity to reduce the number of parameters for estimation,
distributed dictionary learning
\cite{chen15,Chainais,liang14,Chouvardas15,regstap,damdc}, which
employs a bilinear dimensionality-reduction factorization scheme
similar to some low-rank schemes but assumes no regression vectors,
and attribute-distributed learning \cite{Zheng}, which employs
agents and a fusion center to meet communication constraints.
Another important tool in recent related work is the principle of
alternating optimization \cite{Csiszar,Niesen}, which consists of
fixing a set of parameters, adjusting the remaining parameters and
then proceeding in cycles
\cite{Lamare,Lamare1,ling14,bai15,junchi15,magnusson15}. Ling and
Ribeiro have studied dynamic decentralized optimization using the
alternating direction method of multipliers \cite{ling14}. Bai {\it
et al.} have examined alternating optimization procedures to design
sensing matrices and dictionaries for compressive sensing. Yan {\it
et al.} have developed an alternating optimization for multigraph
matching, whereas Magnusson {\it et al.} \cite{magnusson15} have
studied convergence of nonconvex optimization problems.}

In this paper, we propose a scheme for distributed signal processing
and distributed low--rank algorithms for parameter estimation. In
particular, the proposed algorithms are based on an alternating
optimization strategy \cite{Csiszar,Niesen,Lamare,Lamare1} and are
called the distributed reduced-rank joint iterative optimization
normalized least mean squares (DRJIO--NLMS) algorithm and the
distributed reduced-rank joint iterative optimization recursive
least squares (DRJIO--RLS) algorithm. In contrast to prior work on
low-rank techniques \cite{Lamare}-\cite{Wang1} and distributed
methods
\cite{Msechu}-\cite{Chouvardas1,spadf,mfsic,mbdf,tds,armo,badstbc,did,baplnc},
distributed adaptive techniques based on the alternating
optimization strategy are investigated. The proposed low-rank
strategies are distributed and perform dimensionality reduction
without costly decompositions at each agent. The proposed
DRJIO--NLMS and DRJIO--RLS algorithms are flexible with regards to
the amount of information that is exchanged, have low cost and high
performance. We also present a computational complexity analysis of
the proposed and existing low-rank algorithms along with an analysis
of the convergence of the proposed techniques. Applications to
parameter estimation in wireless sensor networks and smart grids are
then studied.

The main contributions of this work can be summarized as:
\begin{itemize}
\item{Distributed low-rank adaptive algorithms based on alternating optimization .}
\item{An analysis of the convergence and the computational complexity of the
proposed distributed algorithms.}
\item{A study of the proposed and existing distributed algorithms in wireless sensor networks and smart
grids.}
\end{itemize}

This paper is organized as follows: In Section II, the system model
and the problem statement are described. In Section III, the
proposed distributed dimensionality reduction and adaptive
processing scheme is presented. Section IV details the proposed
distributed low-rank algorithms. In Section V, an analysis of the
convergence of the proposed algorithms is carried out along with a
study of their computational complexity. Simulation results are
presented and discussed in Section VI, whereas conclusions are drawn
in Section VII.

\section{System Model and Problem Statement}
\begin{figure}[!htb]

\begin{center}
\def\epsfsize#1#2{1\columnwidth}
\epsfbox{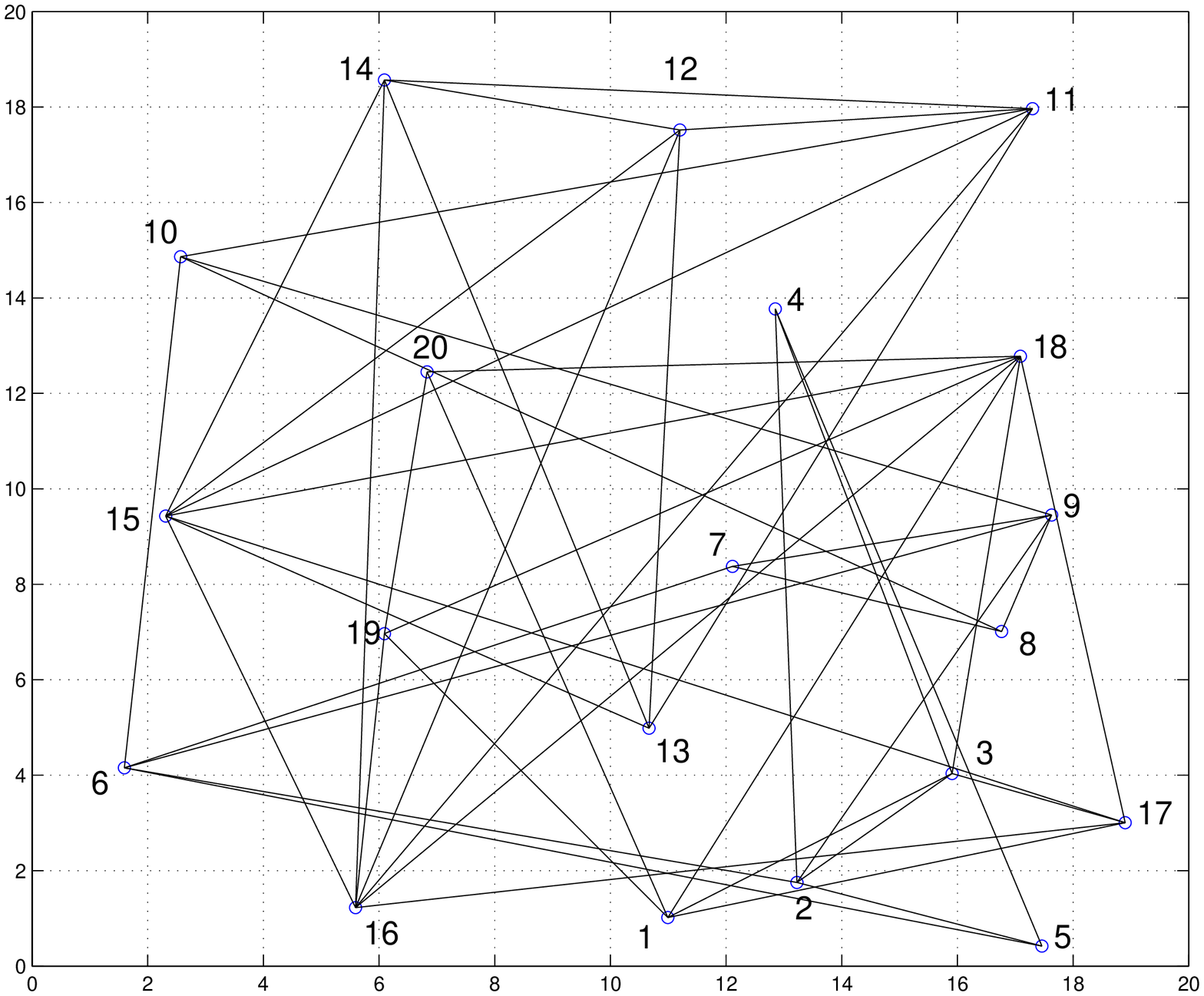}\caption{\footnotesize Network topology with N
nodes} \label{fig6:fig1}
\end{center}
\end{figure}

A distributed network with N nodes, which have limited processing
capabilities, is considered with a partially connected topology as
illustrated in Fig. \ref{fig6:fig1}. A diffusion protocol in which
nodes from the same neighborhood communicate with each other at
every iteration is employed \cite{Lopes2}, although other strategies
such as incremental \cite{Lopes1} and consensus-based \cite{Xie}
could also be used. A partially connected network means that nodes
can exchange information only with their neighbors determined by the
connectivity topology. In contrast, a fully connected network means
that, data broadcast by a node can be captured by all other nodes in
the network \cite{Bertrand}. At every time instant $i$, each node
$k$ takes a scalar measurement $d_k(i)$ according to
\begin{equation}
{d_k(i)} = {\boldsymbol {\omega}}_0^H{\boldsymbol x_k(i)} +{n_k(i)},~~~ \label{Eqn6:desired signal}
i=1,2, \ldots, \textrm{I} ,
\end{equation}
where ${\boldsymbol x_k(i)}$ is the $M \times 1$ input signal vector
with zero mean and variance $\sigma_{x,k}^2$ that is also observed
by node $k$, ${ n_k(i)}$ is the noise sample measured at node $k$
which has zero mean and variance $\sigma_{n,k}^2$. Observing
(\ref{Eqn6:desired signal}), we can see that the measurements for
all nodes are related to an unknown parameter vector ${\boldsymbol
{\omega}}_0$ with size $M \times 1$, that would be estimated by the
network. The aim of such a network is to compute an estimate of
${\boldsymbol{\omega}}_0$ in a distributed fashion, which can
minimize the global cost function
\begin{equation}
{J({\boldsymbol \omega_k}(i))} = \sum_{k=1}^{N}{\mathbb{E}\big|{
d_k(i)}- {\boldsymbol \omega_k}^H(i){\boldsymbol x_k(i)}\big|^2} ,
\end{equation}
where $\mathbb{E}$ denotes expected value and ${\boldsymbol
\omega_k}^H(i)$ is the estimator at time $i$. To solve this problem,
one suitable technique is the adapt--then--combine (ATC) diffusion
strategy \cite{Lopes2} described by
\begin{equation}
\left\{\begin{array}{ll}
{\boldsymbol \psi}_k(i)= {\boldsymbol \omega}_k(i-1)+{\mu}_k {\boldsymbol x_k(i)}\big[{d_k(i)}-{\boldsymbol \omega}_k^H(i-1){\boldsymbol x_k(i)}\big]^*,\\
\ \\
{\boldsymbol {\omega}}_k(i)= \sum\limits_{l\in \mathcal{N}_k} c_{kl} \boldsymbol\psi_l(i),
\end{array}
\right.
\end{equation}
where ${\mu}_k$ is the step size, $\mathcal{N}_k$ indicates the set
of neighbors for node $k$, ${\boldsymbol \psi}_k(i)$ is the local
estimator, $|\mathcal{N}_k|$ denotes the cardinality of
$\mathcal{N}_k$ and $c_{kl}>0$ are the combination coefficients,
which are calculated in this work using the Metropolis rule
\cite{Lopes2} given by
\begin{equation}
\left\{\begin{array}{ll}
c_{kl}= \frac{1}
{max(|\mathcal{N}_k|,|\mathcal{N}_l|)},\ \ \ \ \ \ \ \ \  \ \ \ \ \ \ \ \ \ \ \ $if\  $k\neq l$\  \ are\  linked$\\
c_{kl}=0,              \ \ \ \ \ \ \ \ \ \  \ \ \ \ \ \ \ \ \ \ \ \ \ \ \ \ \ \ \ \ \ \ \ \ \ $for\  $k$\  and\  $l$\ not\  linked$\\
c_{kk} = 1 - \sum\limits_{l\in \mathcal{N}_k / k} c_{kl}, \ \ \ \ \ \ \ \ \ \ \ \ \ \ \ \ \ \ \ $for\  $k$\ =\ $l$$
\end{array}
\right.
\end{equation}
and should satisfy
\begin{equation}
\sum\limits_{l} c_{kl} =1 , l\in \mathcal{N}_k \forall k .
\end{equation}
Note that other combination rules can also be employed. With this
adaptation strategy, when the dimension of the unknown parameter
vector ${\boldsymbol {\omega}}_0$ is large, this could lead to a
high communication overhead between each neighbor node and the
learning speed of the network is reduced. In order to reduce the
communication overhead, accelerate the learning and optimize the
distributed processing, we incorporate at the $k$th node of the
network distributed low-rank strategies based on alternating
optimization techniques.

\section{Distributed Dimensionality Reduction and Adaptive Processing}

\begin{figure}[!htb]
\begin{center}
\def\epsfsize#1#2{1.0\columnwidth}
\epsfbox{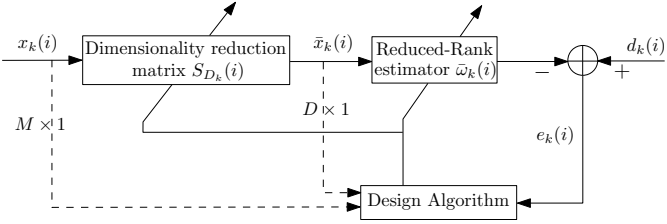}\caption{\footnotesize
Proposed dimensionality reduction scheme at each node or agent}
\label{fig6:fig2}
\end{center}
\end{figure}

The proposed distributed dimensionality reduction scheme, depicted
in Fig.\ref{fig6:fig2}, employs a transformation matrix $\boldsymbol
S_{D_k}(i)$ to process the input signal ${\boldsymbol x_k(i)}$ with
dimensions $M \times 1$ and projects it onto a lower $D \times 1$
dimensional subspace ${\bar{\boldsymbol x}_k(i)}$, where $D\ll M$.
Following this procedure, a low-rank estimator
$\bar{\boldsymbol\omega}_k(i)$ is computed, and the
$\bar{\boldsymbol\omega}_k(i)$ is transmitted by each node. In
particular, the transformation matrix $\boldsymbol S_{D_k}(i)$ and
low-rank estimator $\bar{\boldsymbol\omega}_k(i)$ will be jointly
optimized in the proposed scheme according to the mean squared error
(MSE) criterion.

Specifically, we start the description of the method with an $M\times D$ matrix
$\boldsymbol S_{D_k}(i)$, which carries out a dimensionality reduction on the
input signal of each agent as given by
\begin{equation}
\bar{\boldsymbol x}_k(i) = \boldsymbol S_{D_k}^H(i)\boldsymbol x_k(i),
\end{equation}
where, in what follows, all $D$--dimensional quantities are
designated with an overbar. The design of $\boldsymbol S_{D_k}(i)$
and $\bar{\boldsymbol\omega}_k(i)$ corresponds to the optimization
problem given by
\begin{equation}
\begin{split}
\big\{ \boldsymbol S_{D_k}^{\rm opt}, \bar{\boldsymbol\omega}_k^{\rm
opt} \big\} = \min_{\boldsymbol S_{D_k}(i),
\bar{\boldsymbol\omega}_k(i)}  & \sum_{k=1}^{N}{\mathbb{E} [|{
d_k(i)}- {\boldsymbol{\bar{\omega}}_k}^H(i)\boldsymbol
S_{D_k}^H(i){\boldsymbol x_k(i)}|^2]}\\
& \quad + \delta E[||d_{k}^*(i){\boldsymbol
S}_{D_k}^H(i){\boldsymbol x}_k(i)||^2 \textcolor{red}{]} \\
& \quad + \gamma E[ \sum_{d=1}^{D} {\boldsymbol e}_d^H {\boldsymbol
S}_{D_k}^H(i){\boldsymbol I}_{M,D} {\boldsymbol e}_d ]
\label{Eqn6:new cost}
\end{split}
\end{equation}
where $^{*}$ denotes complex conjugation,
$\boldsymbol{\bar{\omega}}_k(i)$ is the low-rank estimator, the $M
\times 1$ vectors ${\boldsymbol e}_d$ contain one in the $d$th entry
and zeros elsewhere, the $M\times D$ matrix ${\boldsymbol I}_{M,D}$
contains a $D$-dimensional identity matrix on the top and zeros
elsewhere, and the parameters $\gamma$ and $\delta$ are
regularization terms that ensure the solution has rank $D$. In what
follows, we describe the adaptation step that computes the
parameters of $\boldsymbol{\bar{\omega}}_k(i)$ and $\boldsymbol
S_{D_k}(i)$ based on an alternating minimization strategy, which
consists of fixing one set of parameters and then minimizing the
other.

By fixing $\boldsymbol{\bar{\omega}}_k(i)$ and minimizing (\ref{Eqn6:new cost})
with respect to $\boldsymbol S_{D_k}(i)$, we arrive at the following
expression:
\begin{equation}
\boldsymbol S_{D_k}(i)= \boldsymbol R_k^{-1}(i)\boldsymbol
P_{D_k}(i)\bar{\boldsymbol
R}_{\bar{\boldsymbol\omega}_k}^{-1}(i),\label{Eqn6:sdk}
\end{equation}
where the covariance matrix of the input signal vector at node $k$
$\boldsymbol x_k(i)$ is assumed to be full-rank and is given by
\begin{equation}
\boldsymbol R_k(i)=\mathbb{E}[\boldsymbol x_k(i)\boldsymbol x_k^H(i)],
\end{equation}
the cross-correlation matrix is given by
{\begin{equation}
\boldsymbol P_{D_k}(i)=\mathbb{E}[d_k^*(i)(\boldsymbol
x_k(i)\bar{\boldsymbol\omega}_k^H(i)+\gamma {\boldsymbol I}_{M,D})]
\label{Eqn6:pdk}
\end{equation}
and the covariance matrix of the reduced-rank parameter vector is
described by
\begin{equation}
\bar{\boldsymbol
R}_{\bar{\boldsymbol\omega}_k}(i)=\boldsymbol{\bar{\omega}}_k(i)\boldsymbol{\bar{\omega}}_k^H(i)+\delta
{\boldsymbol I}_{D}.
\end{equation}}
We then fix $\boldsymbol S_{D_k}(i)$ and minimize (\ref{Eqn6:new cost}) with
respect to $\boldsymbol{\bar{\omega}}_k(i)$, which results in
\begin{equation}
\boldsymbol{\bar{\omega}}_k(i)= \bar{\boldsymbol R}_k^{-1}(i)\bar{\boldsymbol
p}_k(i),\label{Eqn6:wk}
\end{equation}
where the covariance matrix of the reduced-rank input signal vector
that is also assumed to be full-rank is expressed by
\begin{equation}
\bar{\boldsymbol R}_k(i)=\mathbb{E}[\boldsymbol
S_{D_k}^H(i)\boldsymbol x_k(i)\boldsymbol x_k^H(i)\boldsymbol
S_{D_k}(i)] = \mathbb{E}[\bar{\boldsymbol x}_k(i)\bar{\boldsymbol
x}_k^H(i)]
\end{equation}
and the cross-correlation vector is given by
\begin{equation}
\bar{\boldsymbol p}_k(i)=\mathbb{E}[d_k^*(i)\boldsymbol S_{D_k}^H(i)\boldsymbol
x_k(i)]=\mathbb{E}[d_k^*(i)\bar{\boldsymbol x}_k(i)].
\end{equation}
The reduced-dimension parameter vector
$\boldsymbol{\bar{\omega}}_k(i)$ computed at each agent is then
transmitted as a local low-rank estimator
$\boldsymbol{\bar{\psi}}_k(i)$ to the other agents according to the
network topology. At the receiver of each agent, there is a
combination and reconstruction step in which the received data from
neighboring nodes is combined to obtain a low-rank estimator:
\begin{equation}
\bar{{\boldsymbol {\omega}}}_k(i) = \sum\limits_{l\in \mathcal{N}_k}
c_{kl}\boldsymbol{\bar{\psi}}_l(i),
\end{equation}
The full-dimension estimator is then obtained through a rank-$D$ approximation:
\begin{equation}
{\boldsymbol {\omega}}_k(i)= \boldsymbol S_{D_k}(i){\boldsymbol
{\bar{\omega}}}_k(i),
\end{equation}
which is derived in the Appendix.

The associated low-rank MSE is obtained by substituting the
expressions obtained in (\ref{Eqn6:wk}) and (\ref{Eqn6:sdk}) into
the cost function and is described by \cite{Lamare}
\begin{equation}
{\rm MSE} = \sigma_{d_k}^2-\bar{\boldsymbol
p}_k^H(i)\bar{\boldsymbol R}_k^{-1}(i)\bar{\boldsymbol p}_k(i)
\end{equation}
where $\sigma_{d_k}^2=\mathbb{E}[|d_k(i)|^2]$. Because there is no
closed-form expression for $\boldsymbol S_{D_k}(i)$ and
$\bar{\boldsymbol\omega}_k(i)$ as they depend on each other, a
strategy to compute the parameters is needed. The proposed strategy
is based on an alternating optimization of  $\boldsymbol S_{D_k}(i)$
and $\bar{\boldsymbol\omega}_k(i)$. The rank $D$ must be set by the
designer to ensure appropriate performance taking into account the
bias-variance tradeoff \cite{Scharf}. Furthermore, for the selection
of $D$ the reader is referred to \cite{Qian} for rank selection
methods. In the next section, we develop a distributed low-rank
algorithm to compute the parameters of interest.

\section{Proposed Distributed Low-Rank Algorithms}

In this section, we present the proposed distributed low-rank
adaptive algorithms for distributed estimation, namely DRJIO--NLMS
and DRJIO--RLS. Unlike prior work \cite{Li3,Sayin1,Chouvardas1}, the
proposed algorithms do not require:
\begin{itemize}
\item Additional cost to perform eigen--decompositions \cite{Li3}
\item Extra adaptive processing at the local node \cite{Sayin1}
\item Multiple matrix-vector multiplications to to build the Krylov subspace \cite{Chouvardas1}
\item Costly convex optimization at the local node, which introduces extra complexity \cite{Chouvardas1}.
\end{itemize}
 {The objective of the DRJIO--NLMS and DRJIO--RLS
algorithms is to perform compression/decompression and distributed
parameter estimation subject to the constraint of transmitting only
$D< M$ parameters. The algorithms are flexible, have low cost, very
fast convergence speed and assume that the parameter $D$ is given.
Alternatively, a model-order selection algorithm
\cite{Honig02,Qian03,Lamare2} can be employed to compute $D$ for
each node. In particular, the algorithms rely on an alternating
optimization strategy which consists of fixing a set of parameters
${\boldsymbol S}_{D_k}(i)$, updating the other set of parameters
$\bar{\boldsymbol \omega}_k(i)$, then fixing $\bar{\boldsymbol
\omega}_k(i)$ and updating ${\boldsymbol S}_{D_k}(i)$. This
alternating approach with the recursions for ${\boldsymbol
S}_{D_k}(i)$ and $\bar{\boldsymbol \omega}_k(i)$ is carried out in
cycles until convergence is achieved.}

\subsection{Proposed DRJIO--NLMS algorithm}

In the DRJIO--NLMS algorithm, the parameters in (\ref{Eqn6:new
cost}) are optimized by an alternating procedure that adjusts one of
the parameters while keeping the other parameter fixed using NLMS
recursions.  Therefore, the proposed DRJIO-NLMS algorithm solves the
optimization problem in (\ref{Eqn6:new cost}) in an alternating
fashion.
Using the method of steepest descent, computing the gradient terms
of the cost function in (\ref{Eqn6:new cost}) with respect to
$\boldsymbol S_{D_k}(i)$, replacing the expected value with
instantaneous estimates and considering the recursions in an
alternating fashion, we arrive at the proposed DRJIO--NLMS
algorithm:
\begin{equation}
\begin{split} \boldsymbol S_{D_k}(i) & =\boldsymbol
S_{D_k}(i-1)+\eta(i)e_k^*(i)\boldsymbol
x_k(i)\bar{\boldsymbol\omega}_k^H(i-1)\\
& \quad + \eta(i)\big(\gamma d_k^*(i){\boldsymbol I}_{M,D} - \delta
{\boldsymbol x}_k(i) {\boldsymbol x}_{k}^H(i){\boldsymbol
S}_{D_k}(i-1) \big), \label{Sd_rec}
\end{split}
\end{equation}
\begin{equation}
\bar{\boldsymbol\omega}_k(i)=\bar{\boldsymbol\omega}_k(i-1)+\mu(i)e_k^*(i)\bar{\boldsymbol
x}_k(i),
\end{equation}
where $e_k(i)=d_k(i)-\bar{\boldsymbol\omega}_k^H(i-1)\boldsymbol
S_{D_k}^H(i-1){\boldsymbol x_k(i)}$,
$\mu(i)=\frac{\mu_0}{\boldsymbol x_k^H(i)\boldsymbol x_k(i)}$ and
$\eta(i)=\frac{\eta_0}{\bar{\boldsymbol\omega}_k^H(i-1)\bar{\boldsymbol\omega}_k(i-1)\boldsymbol
x_k^H(i)\boldsymbol x_k(i)}$ are the time--varying step sizes. The
normalization makes the setting of the convergence factors easier,
improves the convergence speed and facilitates the comparison with
other distributed LMS--type algorithms. The recursions are computed
in an alternating way with one iteration per time instant at each
node.

The proposed DRJIO--NLMS algorithm includes two steps, namely, adaptation step
and combination and reconstruction step, which are performed using an
alternating procedure which is detailed next.
\begin{itemize}
\item Adaptation step
\end{itemize}
For the adaptation step, at each time instant $i$=1,2, . . . , I,
each node $k$=1,2, \ldots, N, starts from generating a local
low-rank estimator through
\begin{equation}
{\boldsymbol {\bar{\psi}}}_k(i)=
\bar{\boldsymbol\omega}_k(i-1)+\mu(i)e_k^*(i)\bar{\boldsymbol
x}_k(i),
\end{equation}
where $e_k(i)=d_k(i)-\bar{\boldsymbol\omega}_k^H(i-1)\boldsymbol
S_{D_k}^H(i){\boldsymbol x_k(i)}$. This local low-rank estimator
${\boldsymbol {\bar{\psi}}}_k(i)$ will be transmitted to all its
neighboring nodes under the network topology structure.

Then, each node $k$=1,2, \ldots, N, will locally update its
dimensionality reduction matrix according to (\ref{Sd_rec}) and keep
it locally. Note that $\boldsymbol S_{D_k}(i)$ only employs the
low-rank estimators from neighboring nodes and the local reference
signal $d_k(i)$.

\begin{itemize}
\item Combination and reconstruction step
\end{itemize}
At each time instant $i$=1,2, . . . , I, the combination and
reconstruction step starts after the adaptation step. Each node will
combine the local low-rank estimators from its neighboring nodes and
itself through
\begin{equation}
\bar{{\boldsymbol {\omega}}}_k(i)= \sum\limits_{l\in \mathcal{N}_k} c_{kl}\boldsymbol{\bar{\psi}}_l(i),
\end{equation}
to compute the low-rank estimator $\bar{\boldsymbol {\omega}}_k(i)$.

After the last iteration $I$, each node will reconstruct a full--dimensional
estimator ${\boldsymbol {\omega}}_k(I)$ through the rank-$D$ approximation
given by
\begin{equation}
{\boldsymbol {\omega}}_k(I)= \boldsymbol S_{D_k}(I){\boldsymbol {\bar{\omega}}}_k(I).
\end{equation}
In conclusion, during the distributed processing steps, only the
local low-rank estimator ${\boldsymbol {\bar{\psi}}}_k(i)$ will be
transmitted through the network. The proposed DRJIO--NLMS algorithm
is detailed in Table \ref{table6:table1}.

\begin{table}[!htb]
\centering \caption{{The DRJIO--NLMS Algorithm}}
\begin{tabular}{l}\hline
Initialize: ${\boldsymbol {\bar{\omega}}}_k(0)={\boldsymbol 0}$, ${\boldsymbol S}_{D_k}(0)={\boldsymbol I}_{M,D}$ \\
For each time instant $i$=1,2, . . . , I\\
\ \ \ \ For each node $k$=1,2, \ldots, N\\
\ \ \ \ \ \ \ \ \ \ ${\boldsymbol {\bar{\psi}}}_k(i)= \bar{\boldsymbol\omega}_k(i-1)+\mu(i)e_k^*(i)\bar{\boldsymbol x}_k(i)$\\
\ \ \ \ \ \ \ \ \ \ where $e_k(i)=d_k(i)-\bar{\boldsymbol\omega}_k^H(i-1)\boldsymbol S_{D_k}^H(i){\boldsymbol x_k(i)}$\\
\ \ \ \ \ \ \ \ \ \ \ \% ${\boldsymbol {\bar{\psi}}}_k(i)$ is the local low-rank estimator and will be \\
\ \ \ \ \ \ \ \ \ \ \ \% sent to all neighboring nodes of node $k$ under the network \\
\ \ \ \ \ \ \ \ \ \ \ \%  topology structure.\\
\ \ \ \ \ \ \ \ \ \ $\boldsymbol S_{D_k}(i)=\boldsymbol
S_{D_k}(i-1)+\eta(i)e_k^*(i)\boldsymbol
x_k(i)\bar{\boldsymbol\omega}_k(i-1)$ \\
\ \ \ \ \ \ \ \ \ \ \ \ \ \ \  { $ + \eta(i)\big(\gamma
d_k^*(i){\boldsymbol I}_{M,D} - \delta {\boldsymbol x}_k(i)
{\boldsymbol x}_{k}^H(i){\boldsymbol
S}_{D_k}(i-1) \big)$} \\
\ \ \ \ \ \ \ \ \ \ \% The dimensionality reduction matrix $\boldsymbol S_{D_k}(i)$  \\
\ \ \ \ \ \ \ \ \ \ \% will be updated and kept locally.\\
\ \ \ \ end\\
\ \ \ \ For each node $k$=1,2, \ldots, N\\
\ \ \ \ \ \ \ \ \ \ $\bar{{\boldsymbol {\omega}}}_k(i)= \sum\limits_{l\in \mathcal{N}_k} c_{kl}\boldsymbol{\bar{\psi}}_l(i)$\\
\ \ \ \ \ \ \ \ \ \ \% The low-rank estimator $\bar{{\boldsymbol {\omega}}}_k(i)$ \\
\ \ \ \ \ \ \ \ \ \ \% will be updated and kept locally.\\

\ \ \ \ end\\
end\\
After the last iteration $I$\\
For each node $k$=1,2, \ldots, N\\
\ \ \ \ \% Reconstruction.\\
\ \ \ \ ${\boldsymbol {\omega}}_k(I)= \boldsymbol S_{D_k}(I){\boldsymbol {\bar{\omega}}}_k(I)$ \\
\ \ \ \ where ${\boldsymbol {\omega}}_k(I)$ is the final full--rank estimator.\\
end\\
\hline
\end{tabular}
\label{table6:table1}
\end{table}

\subsection{Proposed DRJIO--RLS algorithm}

 {In this subsection, we develop the DRJIO--RLS
algorithm for computing $\boldsymbol S_{D_k}(i)$ and
$\bar{\boldsymbol\omega}_k(i)$, which is inspired by the derivation
of the standard recursive least squares (RLS) algorithm. The main
differences are that we have two sets of recursions that update the
parameters: one for ${\boldsymbol S}_{D_k}(i)$, which performs
compression/decompression, and another for $\bar{\boldsymbol
\omega}_k(i)$, which performs parameter estimation; and the
recursions are distributed and computed in an alternating fashion.
Therefore, we first fix $\bar{\boldsymbol \omega}_k(i)$ in the
derivation and then derive a set of RLS recursions to compute the
parameters for ${\boldsymbol S}_{D_k}(i)$. Subsequently, we fix
${\boldsymbol S}_{D_k}(i)$ and derive a set of RLS recursions to
compute the parameters for $\bar{\boldsymbol \omega}_k(i)$.}
The DRJIO--RLS algorithm consists of an adaptation step, which computes
$\boldsymbol S_{D_k}(i)$ and $\bar{\boldsymbol\omega}_k(i)$, and a combination
and reconstruction step, which is identical to that of the DRJIO-NLMS
algorithm.
\begin{itemize}
\item{Adaptation step}
\end{itemize}
In order to derive the proposed algorithm, we first define
\begin{equation}
\boldsymbol P_k(i)\triangleq\boldsymbol R_k^{-1}(i),\label{Eqn6:r1}
\end{equation}
\begin{equation}
\boldsymbol P_{D_k}(i)\triangleq\lambda\boldsymbol
P_{D_k}(i-1)+d_k^*(i)\boldsymbol
x_k(i)\bar{\boldsymbol\omega}_k^H(i),
\end{equation}
\begin{equation}
\boldsymbol
Q_{\bar{\boldsymbol\omega}_k}(i)\triangleq\bar{\boldsymbol
R}_{\bar{\boldsymbol\omega}_k}^{-1}(i-1),
\end{equation}
and rewrite the expression in (\ref{Eqn6:sdk}) as follows
\begin{equation}
\begin{split}
\boldsymbol S_{D_k}(i)&= \boldsymbol R_k^{-1}(i)\boldsymbol P_{D_k}(i)\bar{\boldsymbol R}_{\bar{\boldsymbol\omega}_k}^{-1}(i-1)\\
&=\boldsymbol P_k(i)\boldsymbol P_{D_k}(i)\boldsymbol Q_{\bar{\boldsymbol\omega}_k}(i)\\
&=\lambda\boldsymbol P_k(i)\boldsymbol P_{D_k}(i-1)\boldsymbol Q_{\bar{\boldsymbol\omega}_k}(i)+d_k^*(i)\boldsymbol P_k(i)\boldsymbol x_k(i)\bar{\boldsymbol\omega}_k^H(i)\boldsymbol Q_{\bar{\boldsymbol\omega}_k}(i)\\
&=\boldsymbol S_{D_k}(i-1)+\boldsymbol k_k(i)\bigg[d_k^*(i)\boldsymbol t_k^H(i)-\boldsymbol x_k^H(i)\boldsymbol S_{D_k}(i-1)\bigg],
\end{split}
\end{equation}
where the $D\times1$ vector $\boldsymbol t_k(i)=\boldsymbol Q_{\bar{\boldsymbol\omega}_k}(i)\bar{\boldsymbol\omega}_k(i)$ and
the $M\times1$ Kalman gain vector is
\begin{equation}
\boldsymbol k_k(i)=\frac{\lambda^{-1}\boldsymbol P_k(i-1)\boldsymbol x_k(i)}{1+\lambda^{-1}\boldsymbol x_k^H(i)\boldsymbol P_k(i-1)\boldsymbol x_k(i)}.
\end{equation}
In addition, the update for the $M\times M$ matrix $\boldsymbol P_k(i)$ employs the matrix inversion lemma \cite{Haykin} as follows:
\begin{equation}
\boldsymbol P_k(i)=\lambda^{-1}\boldsymbol P_k(i-1)-\lambda^{-1}\boldsymbol k_k(i)\boldsymbol x_k^H(i)\boldsymbol P_k(i-1)
\end{equation}
and the $D\times1$ vector $\boldsymbol t_k(i)$ is updated as
\begin{equation}
\boldsymbol t_k(i)=\frac{\lambda^{-1}\boldsymbol Q_{\bar{\boldsymbol\omega}_k}(i-1)\bar{\boldsymbol\omega}_k(i-1)}{1+\lambda^{-1}\bar{\boldsymbol\omega}_k^H(i-1)\boldsymbol Q_{\bar{\boldsymbol\omega}_k}(i-1)\bar{\boldsymbol\omega}_k(i-1)}.
\end{equation}
The matrix inversion lemma \cite{Haykin} is then used to update the $D\times D$ matrix $\boldsymbol Q_{\bar{\boldsymbol\omega}_k}(i)$ as described by
\begin{equation}
\boldsymbol Q_{\bar{\boldsymbol\omega}_k}(i)=\lambda^{-1}\boldsymbol Q_{\bar{\boldsymbol\omega}_k}(i-1)-\lambda^{-1}\boldsymbol t_k(i)\bar{\boldsymbol\omega}_k^H(i-1)\boldsymbol Q_{\bar{\boldsymbol\omega}_k}(i-1).\label{Eqn6:r2}
\end{equation}
Equations (\ref{Eqn6:r1})--(\ref{Eqn6:r2}) constitute the key steps of the
proposed DRJIO-RLS algorithm for computing $\boldsymbol S_{D_k}(i)$.

To derive the expression for updating $\bar{\boldsymbol\omega}_k(i)$, the
following associated quantities are defined
\begin{equation}
\bar{\boldsymbol \Phi}_k(i)\triangleq\bar{\boldsymbol R}_k^{-1}(i)\label{Eqn6:r3}
\end{equation}
\begin{equation}
\bar{\boldsymbol p}_k(i)=\lambda\bar{\boldsymbol p}_k(i-1)+d_k^*(i)\boldsymbol x_k(i).
\end{equation}
Then, equation (\ref{Eqn6:wk}) will be rewritten as
\begin{equation}
\begin{split}
\boldsymbol{\bar{\omega}}_k(i)&= \bar{\boldsymbol R}_k^{-1}(i)\bar{\boldsymbol p}_k(i)\\
&=\bar{\boldsymbol \Phi}_k(i)\bar{\boldsymbol p}_k(i)\\
&=\lambda\bar{\boldsymbol \Phi}_k(i)\bar{\boldsymbol p}_k(i-1)+d_k^*(i)\bar{\boldsymbol \Phi}_k(i)\boldsymbol x_k(i)\\
&=\boldsymbol{\bar{\omega}}_k(i-1)+\bar{\boldsymbol k}_k(i)\bigg[d_k^*(i)-\bar{\boldsymbol x}_k^H(i)\boldsymbol{\bar{\omega}}_k(i-1)\bigg],
\end{split}
\end{equation}
where the $D\times1$ Kalman gain vector is given by
\begin{equation}
\bar{\boldsymbol k}_k(i)=\frac{\lambda^{-1}\bar{\boldsymbol \Phi}_k(i-1)\bar{\boldsymbol x}_k(i)}{1+\lambda^{-1}\bar{\boldsymbol x}_k^H(i)\bar{\boldsymbol \Phi}_k(i-1)\bar{\boldsymbol x}_k(i)}.
\end{equation}
and the update for the matrix inverse $\bar{\boldsymbol \Phi}_k(i)$ employs the matrix inversion lemma \cite{Haykin}
\begin{equation}
\bar{\boldsymbol \Phi}_k(i)=\lambda^{-1}\bar{\boldsymbol \Phi}_k(i-1)-\lambda^{-1}\bar{\boldsymbol k}_k(i)\bar{\boldsymbol x}_k^H(i)\bar{\boldsymbol \Phi}_k(i-1).\label{Eqn6:r4}
\end{equation}
Equations (\ref{Eqn6:r3})--(\ref{Eqn6:r4}) are the key steps of the
proposed DRJIO-RLS algorithm for computing
$\boldsymbol{\bar{\omega}}_k(i)$. Since the combination and
reconstruction step is identical to that of DRJIO-NLMS we omit it
here. The proposed DRJIO--RLS algorithm is detailed in Table
\ref{table6:table2}.

\begin{table}[!htb]
\centering \caption{The DRJIO-RLS Algorithm}
\begin{tabular}{l}\hline
Initialize: ${\boldsymbol {\bar{\omega}}}_k(0)$=0\\
\ \ \ \ \ \ \ \ \ \ \ \ $\boldsymbol P_k(0)=\delta^{-1}\boldsymbol I_{M\times M}$, $\boldsymbol Q_{\bar{\boldsymbol\omega}_k}(0)=\delta^{-1}\boldsymbol I_{D\times D}$,\\
\ \ \ \ \ \ \ \ \ \ \ \ $\bar{\boldsymbol \Phi}_k(0)=\delta^{-1}\boldsymbol I_{D\times D}$ and $\delta=$ small positive constant\\
For each time instant $i$=1,2, . . . , I\\
\ \ \ \ For each node $k$=1,2, \ldots, N\\
\ \ \ \ \ \ \ \ \ \ $\boldsymbol k_k(i)=\frac{\lambda^{-1}\boldsymbol P_k(i-1)\boldsymbol x_k(i)}{1+\lambda^{-1}\boldsymbol x_k^H(i)\boldsymbol P_k(i-1)\boldsymbol x_k(i)}$ \\
\ \ \ \ \ \ \ \ \ \ $\boldsymbol t_k(i)=\frac{\lambda^{-1}\boldsymbol Q_{\bar{\boldsymbol\omega}_k}(i-1)\bar{\boldsymbol\omega}_k(i-1)}{1+\lambda^{-1}\bar{\boldsymbol\omega}_k^H(i-1)\boldsymbol Q_{\bar{\boldsymbol\omega}_k}(i-1)\bar{\boldsymbol\omega}_k(i-1)}$\\
\ \ \ \ \ \ \ \ \ \ $\boldsymbol S_{D_k}(i)=\boldsymbol S_{D_k}(i-1)+\boldsymbol k_k(i)\bigg[d_k^*(i)\boldsymbol t_k^H(i)-\boldsymbol x_k^H(i)\boldsymbol S_{D_k}(i-1)\bigg]$\\
\ \ \ \ \ \ \ \ \ \ $\boldsymbol P_k(i)=\lambda^{-1}\boldsymbol P_k(i-1)-\lambda^{-1}\boldsymbol k_k(i)\boldsymbol x_k^H(i)\boldsymbol P_k(i-1)$\\
\ \ \ \ \ \ \ \ \ \ $\boldsymbol Q_{\bar{\boldsymbol\omega}_k}(i)=\lambda^{-1}\boldsymbol Q_{\bar{\boldsymbol\omega}_k}(i-1)-\lambda^{-1}\boldsymbol t_k(i)\bar{\boldsymbol\omega}_k^H(i-1)\boldsymbol Q_{\bar{\boldsymbol\omega}_k}(i-1)$\\
\ \ \ \ \ \ \ \ \ \ $\bar{\boldsymbol k}_k(i)=\frac{\lambda^{-1}\bar{\boldsymbol \Phi}_k(i-1)\bar{\boldsymbol x}_k(i)}{1+\lambda^{-1}\bar{\boldsymbol x}_k^H(i)\bar{\boldsymbol \Phi}_k(i-1)\bar{\boldsymbol x}_k(i)}$\\
\ \ \ \ \ \ \ \ \ \ $\boldsymbol{\bar{\psi}}_k(i)=\boldsymbol{\bar{\omega}}_k(i-1)+\bar{\boldsymbol k}_k(i)\bigg[d_k^*(i)-\bar{\boldsymbol x}_k^H(i)\boldsymbol{\bar{\omega}}_k(i-1)\bigg]$\\
\ \ \ \ \ \ \ \ \ \ $\bar{\boldsymbol \Phi}_k(i)=\lambda^{-1}\bar{\boldsymbol \Phi}_k(i-1)-\lambda^{-1}\bar{\boldsymbol k}_k(i)\bar{\boldsymbol x}_k^H(i)\bar{\boldsymbol \Phi}_k(i-1)$\\
\ \ \ \ end\\
\ \ \ \ For each node $k$=1,2, \ldots, N\\
\ \ \ \ \ \ \ \ \ \ $\bar{{\boldsymbol {\omega}}}_k(i)= \sum\limits_{l\in \mathcal{N}_k} c_{kl}\boldsymbol{\bar{\psi}}_l(i)$\\
\ \ \ \ end\\
end\\
For each node $k$=1,2, \ldots, N\\
\ \ \ \ \% Reconstruction.\\
\ \ \ \ ${\boldsymbol {\omega}}_k(I)= \boldsymbol S_{D_k}(I){\boldsymbol {\bar{\omega}}}_k(I)$ \\
\ \ \ \ where ${\boldsymbol {\omega}}_k(I)$ is the final full--rank estimator.\\
end\\
\hline
\end{tabular}
\label{table6:table2}
\end{table}

\section{Analysis of the proposed algorithms}

In this section, the computational complexity of the proposed
algorithms is detailed and an analysis of sufficient conditions for
convergence and a convergence proof to the optimal low-rank
estimator are developed.
 {Regarding our convergence proof, it is worth noting that the
proof described in \cite{Csiszar} was performed for a non-adaptive scenario,
whereas the work in \cite{Niesen} was carried out for an adaptive setting. Our
proof has been developed for the distributed adaptive case. }

\subsection{Computational Complexity Analysis}

Here, we evaluate the computational complexity of the proposed
DRJIO--NLMS and DRJIO--RLS algorithms. The computational complexity
of the proposed DRJIO--NLMS algorithm is $O(DM)$, while the proposed
DRJIO--RLS algorithm has a complexity $O(M^2+D^2)$, where $O(\cdot)$
is used to classify algorithms according to how their requirements
in arithmetic operations grow as the input size grows. The
distributed NLMS algorithm \cite{Lopes2} requires $O(M)$, while the
complexity of the distributed RLS algorithm \cite{Cattivelli2} is
$O(M^2)$. For the Krylov Subspace NLMS \cite{Chouvardas1} the
complexity reaches $O(DM^2)$, while for the distributed principal
subspace estimation algorithms \cite{Li3}, the complexity is
$O(M^3)$. Thus, the proposed DRJIO--NLMS algorithm has a much lower
computational complexity, and because we consider $D\ll M$, it has a
comparable cost to the distributed NLMS algorithm \cite{Lopes2}. The
computational complexity of the model-order selection algorithm of
\cite{Lamare2} with extended filters requires $3(D_{\rm max} -D_{\rm
min}) +1$ additions and a sorting algorithm to find the best model
order. An additional and very important aspect of distributed
low-rank algorithms is that the dimensionality reduction results in
a decrease in the number of transmitted parameters from $M$ to $D$,
which corresponds to a less stringent bandwidth requirement.

\begin{figure}
\begin{center}
\def\epsfsize#1#2{1.0\columnwidth}
\epsfbox{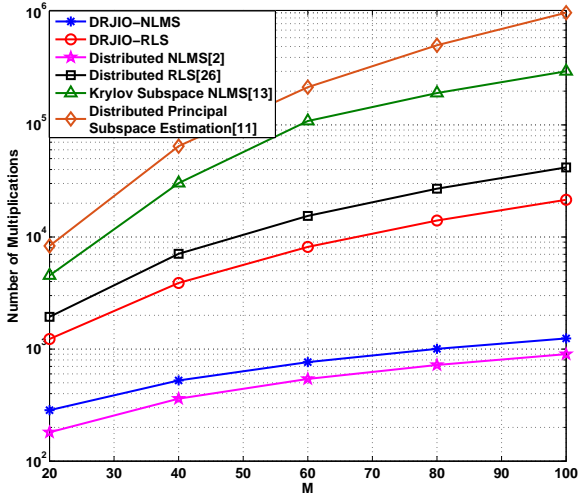}\caption{\footnotesize Complexity in terms of
multiplications}  \label{fig6:cc}
\end{center}
\end{figure}

The details of the computational complexity of the the proposed and
the existing algorithms, are shown in Table \ref{table6:cc}, where
$M$ is the length of the unknown parameter that needs to be
estimated, $D$ is the reduced dimension and $|\mathcal{N}_k|$ is the
cardinality of $\mathcal{N}_k$. To further illustrate the
computational complexity for different algorithms, we present the
main trends in terms of the number of multiplications for the
proposed and existing algorithms in Fig. \ref{fig6:cc}. For the
parameters, we consider a network with $N=20$ nodes, take node 14 as
an example, and set $D=5$ and $|\mathcal{N}_k|=5$.

\begin{table*}
\centering
\caption{Computational Complexity of Different Algorithms}
\begin{tabular}{|c|c|c|}
\hline
Algorithm&Multiplications&Additions\\
\hline
DRJIO--NLMS&$2(D+1)M+(3+|\mathcal{N}_k|)D+5$&$(2D+1)M+(2+|\mathcal{N}_k|)D-2$\\
\hline
DRJIO--RLS&$2M^2+(3+2D)M+4D^2$&$2M^2+2DM+4D^2$\\
&$+(9+|\mathcal{N}_k|)D$&$+(2+|\mathcal{N}_k|)D$\\
\hline
Distributed NLMS \cite{Lopes2}&$(4+|\mathcal{N}_k|)M+1$&$(5+|\mathcal{N}_k|)M-1$\\
\hline
Distributed RLS \cite{Cattivelli2}&$4M^2+(12+|\mathcal{N}_k|)M-1$&$4M^2+(16+|\mathcal{N}_k|)M+1$\\
\hline
Krylov Subspace&$6DM^2+4M+(5+|\mathcal{N}_k|)D$&$6DM^2+2M+(2+|\mathcal{N}_k|)D$\\
NLMS \cite{Chouvardas1}&&\\
\hline
Distributed principal&$M^3+2(D+2)M$&$M^3+(D+1)M$\\
subspace estimation \cite{Li3}&$+(3+|\mathcal{N}_k|)D+4$&$+(2+|\mathcal{N}_k|)D-1$\\
\hline
\end{tabular}
\label{table6:cc}
\end{table*}

\subsection{Sufficient Conditions for Convergence}

 {To start the analysis, we assume that the transformation matrix
$\boldsymbol S_{D_k}(i)$ that performs compression and
decompression/reconstruction and the low-rank parameter estimator
${\boldsymbol {\omega}}_k(i)$ aim to estimate the optimum pair
$\boldsymbol S_{D_{k,{\rm opt}}}$ and $\bar{\boldsymbol
{\omega}}_{k,{\rm opt}}(i)$ containing a common set of parameters of
interest in the network}. Then, to develop the analysis and proofs,
we need to define a metric space and the Hausdorff distance that
will extensively be used. A metric space is an ordered pair
$(\mathcal{M}, r)$, where $\mathcal{M}$ is a nonempty set, and $r$
is a metric on $\mathcal{M}$, i.e., a function
$r:\mathcal{M}\times\mathcal{M}\rightarrow\mathbb{R}$ such that, for
any $x, y, z$, and $\mathcal{M}$, the following conditions hold.
\begin{itemize}
\item[1)] $r(x,y)\geq0$.
\item[2)] $r(x,y)=0$ if $x=y$.
\item[3)] $r(x,y)=r(y,x)$.
\item[4)] $r(x,y)\leq r(x,y) + r(y,z)$  (triangle inequality).
\end{itemize}

The Hausdorff distance measures how far two subsets of a metric space are from each other and is defined by
\begin{equation}
r_H(X,Y)=\max\bigg\{\sup_{x\in X}\inf_{y\in Y} r(x,y), \sup_{y\in Y}\inf_{x\in X} r(x,y)\bigg\}.
\end{equation}

 {The proposed algorithms can be stated as alternating
minimization strategies performed in a distributed fashion and
expressed as
\begin{equation}
\boldsymbol S_{D_k}(i)\in\arg\min_{\boldsymbol S_{D_k}^{\rm
opt}\in\underline{\boldsymbol S}_{D_k}(i)} {D}_k\bigg(\boldsymbol
S_{D_k}^{\rm opt},\bar{\boldsymbol\omega}_k(i)\bigg)\ \ \ \
\textrm{for}\  k=1,2, \ldots, N,
\end{equation}
where ${D}_k (\cdot)$ is a distance metric and
\begin{equation}
\bar{\boldsymbol\omega}_k(i)\in\arg\min_{\bar{\boldsymbol\omega}_k^{\rm
opt}\in\bar{\underline{\boldsymbol\omega}}_k(i)}{D}_k\bigg(\boldsymbol
S_{D_k}(i),\bar{\boldsymbol\omega}_k^{\rm opt}\bigg)\ \ \ \
\textrm{for}\  k=1,2, \ldots, N
\end{equation}
where $\boldsymbol S_{D_k}^{\rm opt}$ and
$\bar{\boldsymbol\omega}_k^{\rm opt}$ correspond to the optimal
values of $\boldsymbol S_{D_k}(i)$ and
$\bar{\boldsymbol\omega}_k(i)$, respectively, and the sequences of
compact sets $\{\underline{\boldsymbol S}_{D_k}(i)\}_{i\geq 0}$ and
$\{\bar{\underline{\boldsymbol\omega}}_k(i)\}_{i\geq 0}$ converge to
the sets $\underline{\boldsymbol S}_{D_k,{\rm opt}}$ and
$\underline{\bar{\boldsymbol\omega}}_{k,{\rm opt}}$, respectively.}

 {The sets $\underline{\boldsymbol S}_{D_k,{\rm opt}}$
and $\underline{\bar{\boldsymbol\omega}}_{k,{\rm opt}}$ are not
directly given, but we observe the sequence of compact sets
$\{\underline{\boldsymbol S}_{D_k}(i)\}_{i\geq 0}$ and
$\{\bar{\underline{\boldsymbol\omega}}_k(i)\}_{i\geq 0}$. The goal
of the proposed algorithms is to find a sequence of $\boldsymbol
S_{D_k}(i)$ and $\bar{\boldsymbol\omega}_k(i)$ in a distributed way
such that
\begin{equation}
\lim_{i\rightarrow\infty} {D}_k \bigg(\boldsymbol S_{D_k}(i),
\bar{\boldsymbol\omega}_k(i)\bigg)= {D}_k\bigg(\boldsymbol
S_{D_k}^{\rm opt}, \bar{\boldsymbol\omega}_k^{\rm opt}\bigg).
\end{equation}}
 {To present a set of sufficient conditions under
which the proposed algorithms converge, we employ the so--called
three-- and four--point properties \cite{Csiszar,Niesen,vandevel},
which are used in the study of the theory of convex sets.}
 {Let us assume that there is a function $f :
\mathcal{M}\times \mathcal{M}\rightarrow\mathbb{R}$ such that the
following conditions are satisfied.}
\begin{itemize}
\item[1)]  {\emph{Three--point property} ($\boldsymbol S_{D_k}^{\rm opt}$, $\tilde{\boldsymbol S}_{D_k}$,
$\bar{\boldsymbol\omega}_k^{\rm opt} $). For all $i\geq1$,
$\boldsymbol S_{D_k}^{\rm opt}\in\underline{\boldsymbol
S}_{D_k}(i)$, $\bar{\boldsymbol\omega}_k^{\rm opt}\in
\underline{\bar{\boldsymbol\omega}}_k(i)$ and $\tilde{\boldsymbol
S}_{D_k}\in \arg\min_{\bar{\boldsymbol\omega}_k^{\rm opt}\in
\underline{\bar{\boldsymbol\omega}}_k(i)}{D}_k\bigg(\boldsymbol
S_{D_k}^{\rm opt}, \bar{\boldsymbol\omega}_k^{\rm opt}\bigg)$, we
have}
\begin{equation}
 {f \bigg(\boldsymbol S_{D_k}^{\rm opt},
\tilde{\boldsymbol S}_{D_k}\bigg)+{D}_k\bigg(\tilde{\boldsymbol
S}_{D_k}, \bar{\boldsymbol\omega}_k^{\rm opt} \bigg)\leq
{D}_k\bigg(\boldsymbol S_{D_k}^{\rm opt},
\bar{\boldsymbol\omega}_k^{\rm opt} \bigg).}
\end{equation}
\item[2)]  {\emph{Four--point property}
($\boldsymbol S_{D_k}^{\rm opt}$, $\bar{\boldsymbol\omega}_k^{\rm
opt} $, $\tilde{\boldsymbol S}_{D_k}$,
$\tilde{\bar{\boldsymbol\omega}}_k$). For all $i\geq1$, $\boldsymbol
S_{D_k}^{\rm opt}$, $\tilde{\boldsymbol
S}_{D_k}\in\underline{\boldsymbol S}_{D_k}(i)$,
$\bar{\boldsymbol\omega}_k^{\rm opt}\in
\underline{\bar{\boldsymbol\omega}}_k(i)$ and
$\tilde{\bar{\boldsymbol\omega}}_k\in
\arg\min_{\bar{\boldsymbol\omega}_k^{\rm opt}\in
\underline{\bar{\boldsymbol\omega}}_k(i)} {D}_k
\bigg(\tilde{\boldsymbol S}_{D_k}, \bar{\boldsymbol\omega}_k^{\rm
opt}\bigg)$, we have}
\begin{equation}
 {{D}_k \bigg(\boldsymbol S_{D_k}^{\rm opt},
\tilde{\bar{\boldsymbol\omega}}_k\bigg)\leq {D}_k\bigg(\boldsymbol
S_{D_k}^{\rm opt}, \bar{\boldsymbol\omega}_k^{\rm opt}
\bigg)+f\bigg(\boldsymbol S_{D_k}^{\rm opt}, \tilde{\boldsymbol
S}_{D_k}\bigg).}
\end{equation}
\end{itemize}

 {\emph{Theorem}: Let $\{\underline{\boldsymbol
S}_{D_k}(i)\}_{i\geq 0}$,
$\{\bar{\underline{\boldsymbol\omega}}_k(i)\}_{i\geq 0}$,
$\underline{\boldsymbol S}_{D_k}^{\rm opt}$,
$\underline{\bar{\boldsymbol\omega}}_k^{\rm opt}$ be compact subsets
of the compact metric space $(\mathcal{M}, r)$ such that}
 {\begin{equation} \underline{\boldsymbol
S}_{D_k}(i)\stackrel{r_h}{\rightarrow} \underline{\boldsymbol
S}_{D_k}^{\rm opt} \ \ \ \
\bar{\underline{\boldsymbol\omega}}_k(i)\stackrel{r_h}{\rightarrow}\underline{\bar{\boldsymbol\omega}}_k^{\rm
opt}
\end{equation}
and let ${D}_k: \mathcal{M}\times\mathcal{M}\rightarrow\mathbb{R}$
be a continuous function.}

 {Now, let conditions 1) and 2) hold. Then, for the
proposed algorithms, we have
\begin{equation}
\lim_{i\rightarrow\infty} {D}_k \bigg(\boldsymbol S_{D_k}(i),
\bar{\boldsymbol\omega}_k(i)\bigg)= {D}_k \bigg(\boldsymbol
S_{D_k}^{\rm opt}, \bar{\boldsymbol\omega}_k^{\rm opt}\bigg).
\end{equation}
A general proof of this theorem is detailed in
\cite{Csiszar,Niesen}.}

\subsection{Convergence to the Optimal Low-Rank Estimator}

In this section, we show that the proposed low--rank algorithm
globally and exponentially converges to the optimal low-rank
estimator \cite{Scharf,Hua}. This result is applicable to
least-squares type algorithms with forgetting factor $\lambda=1$. We
remark that for stochastic gradient (or LMS) algorithms and
least-squares algorithms with forgetting factor $\lambda\neq 1$,
there will be a misadjustment or loss in MSE due to the adaptation
with the step size when an LMS algorithms is adopted, or due to the
forgetting factor when an RLS algorithm is chosen. To proceed with
our proof, let us rewrite the expressions in (\ref{Eqn6:sdk}) and
(\ref{Eqn6:pdk} for time instant zero as follows:
\begin{equation}
\boldsymbol R_k(0)\boldsymbol S_{D_k}(0)\bar{\boldsymbol
R}_{\bar{\boldsymbol\omega}_k}(0)=\boldsymbol P_{D_k}(0)=\boldsymbol
p_k(0)\boldsymbol{\bar{\omega}}_k^H(0) + \delta {\boldsymbol
\Upsilon}, \label{n1}
\end{equation}
where ${\boldsymbol \Upsilon} = \Big[\begin{array}{c}
                                  {\boldsymbol I}_D \\
                                  {\boldsymbol 0}_{M-D \times D}
                                \end{array}\Big]$,
${\boldsymbol \Upsilon}$ is an $M\times D$ matrix containing an
identity matrix ${\boldsymbol I}_D$ with size $D$ and an $M-D \times
D$ matrix with zeros ${\boldsymbol 0}_{M-D \times D}$, $\delta$ is a
small positive scalar used to regularize the recursion at
initialization and ensure that a rank-$D$ matrix $\boldsymbol
P_{D_k}(0)$ is obtained, and the $D$-dimensional set of normal
equations that must be solved to compute
$\boldsymbol{\bar{\omega}}_k(1)$ is given by
\begin{equation}
\begin{split}
\bar{\boldsymbol R}_k(0)\boldsymbol{\bar{\omega}}_k(1)& =\boldsymbol
S_{D_k}^H(0) \boldsymbol R_k(0)\boldsymbol
S_{D_k}(0)\boldsymbol{\bar{\omega}}_k(1)\\ & =\boldsymbol
S_{D_k}(0){\boldsymbol p}_k(0)= \bar{\boldsymbol p}_k(0),\label{n2}
\end{split}
\end{equation}
where ${\boldsymbol p}_k(i) =E[d_k(i)^* {\boldsymbol x}_k(i)$ is the
cross-correlation vector.

Using (\ref{n1}), we can obtain the following relation
 {
\begin{align}
\bar{\boldsymbol R}_{\bar{\boldsymbol\omega}_k}(0)&=\bigg(\boldsymbol S_{D_k}^H(0)\boldsymbol R_k^2(0)\boldsymbol S_{D_k}(0)\bigg)^{-1}\notag\\
&\times\boldsymbol S_{D_k}(0)\boldsymbol R_k(0) \boldsymbol
P_{D_k}(0) .
\end{align}
Substituting the aforementioned result for $\bar{\boldsymbol
R}_{\bar{\boldsymbol\omega}_k}(0)$ into the expression in
(\ref{n1}), we get a recursive expression for $\boldsymbol
S_{D_k}(0)$ as
\begin{align}
\boldsymbol S_{D_k}(0)&=\boldsymbol R_k^{-1}(0) \boldsymbol P_{D_k}(0)\bigg(\boldsymbol S_{D_k}^H(0)\boldsymbol R_k(0)
\boldsymbol P_{D_k}(0)\bigg)^{-1}\notag\\
&\times\bigg(\boldsymbol S_{D_k}^H(0)\boldsymbol R_k^2(0)\boldsymbol
S_{D_k}(0)\bigg)^{-1}.
\end{align}
Using (\ref{n1}), we can express $\boldsymbol{\bar{\omega}}_k(1)$ as
\begin{equation}
\boldsymbol{\bar{\omega}}_k(1)=\bigg(\boldsymbol S_{D_k}^H(0)\boldsymbol R_k(0)\boldsymbol S_{D_k}(0)\bigg)^{-1}\boldsymbol S_{D_k}^H(0)\boldsymbol p_k(0).
\end{equation}
For the proposed DRJIO--NLMS and DRJIO--RLS, the relation is given
by
\begin{equation}
\boldsymbol{\omega}_k(1)=\boldsymbol S_{D_k}(1)\sum_{l\in
\mathcal{N}_k}c_{kl}\boldsymbol{\bar{\omega}}_l(1). \label{eq60}
\end{equation}
Substituting $\boldsymbol S_{D_k}(1)$ and
$\boldsymbol{\bar{\omega}}_l(1)$ into (\ref{eq60}), we obtain
\begin{align}
\boldsymbol{\omega}_k(1)&=\boldsymbol R_k^{-1}(1)\boldsymbol P_{D_k}(1)\bigg(\boldsymbol S_{D_k}^H(1)
\boldsymbol R_k(1)\boldsymbol P_{D_k}(1)\bigg)^{-1}\notag\\
&\quad \times\bigg(\boldsymbol S_{D_k}^H(1)\boldsymbol R_k^2(1)\boldsymbol S_{D_k}(1)\bigg)^{-1}\notag\\
&\quad \times \sum_{l\in \mathcal{N}_k}c_{kl}\bigg(\boldsymbol
S_{D_l}^H(0)\boldsymbol R_l(0)\boldsymbol S_{D_l}(0)\bigg)^{-1}\boldsymbol
S_{D_l}^H(0)\boldsymbol p_l(0).
\end{align}
More generally, we can express the proposed distributed algorithms by the
following recursion:
\begin{align}
\boldsymbol{\omega}_k(i)&=\boldsymbol S_{D_k}(i)\sum_{l\in \mathcal{N}_k}c_{kl}\boldsymbol{\bar{\omega}}_l(i)\notag\\
&=\boldsymbol R_k^{-1}(i)\boldsymbol P_{D_k}(i)\bigg(\boldsymbol S_{D_k}^H(i)\boldsymbol R_k(i)\boldsymbol P_{D_k}(i)\bigg)^{-1}\notag\\
&\quad \times\bigg(\boldsymbol S_{D_k}^H(i)\boldsymbol R_k^2(i)\boldsymbol S_{D_k}(i)\bigg)^{-1}\notag\\
&\quad \times \sum_{l\in \mathcal{N}_k}c_{kl}\bigg(\boldsymbol S_{D_l}^H(i-1)\boldsymbol R_l(i-1)\boldsymbol S_{D_l}(i-1)\bigg)^{-1}\notag\\
&\quad \times\boldsymbol S_{D_l}^H(i-1)\boldsymbol p_l(i-1).\label{wk3}
\end{align}}
 {At this point, we resort to the assumption that the
matrices $\boldsymbol S_{D_k}(i)$ for each node $k$ must converge to
the same values, which correspond to the optimal transformation
matrix}. Because the optimal low-rank filter can be described by the
eigenvalue decomposition of $\boldsymbol R_k^{-1/2}(i)\boldsymbol
p_k(i)$ [20], [21], where $\boldsymbol R_k^{-1/2}(i)$ is the square
root of the matrix $\boldsymbol R_k(i)$, and $\boldsymbol p_k(i)$ is
the cross-correlation vector, we have
\begin{equation}
\boldsymbol R_k^{-1/2}(i)\boldsymbol p_k(i)=\boldsymbol\Phi_k\boldsymbol\Lambda_k\boldsymbol\Phi_k^H\boldsymbol p_k(i),\label{wk4}
\end{equation}
where $\boldsymbol\Lambda_k$ is an $M \times M$ diagonal matrix with the
eigenvalues of $\boldsymbol R_k$, and $\boldsymbol\Phi_k$ is a $M \times M$
unitary matrix with the eigenvectors of $\boldsymbol R_k$.

Let us also assume that there exists some $\boldsymbol{\omega}_k(0)$ such that
the randomly selected $\boldsymbol S_{D_k}(0)$ can be written as [21]
\begin{equation}
\boldsymbol S_{D_k}(0)=\boldsymbol R_k^{-1/2}(i)\boldsymbol\Phi_k\boldsymbol{\omega}_k(0),\label{wk5}
\end{equation}
Using (\ref{wk4}) and (\ref{wk5}) in (\ref{wk3}) together with the assumption
and some manipulation of the algebraic expressions, we can express (\ref{wk3})
in a more compact way that is suitable for analysis, as given by
\begin{align}
\boldsymbol{\omega}_k(i)&=\sum_{l\in \mathcal{N}_k}c_{kl}\boldsymbol\Lambda_l^2\boldsymbol{\omega}_l(i-1)\big(\boldsymbol{\omega}_l^H(i-1)\boldsymbol\Lambda_l^2\boldsymbol{\omega}_l(i-1)\big)^{-1}\notag\\
& \quad \times\boldsymbol\omega_l^H(i-1)\boldsymbol\omega_l(i-1). \label{wk2}
\end{align}
The aforementioned expression can be decomposed as follows:
\begin{equation}
\boldsymbol{\omega}_k(i)=\sum_{l\in\mathcal{N}_k}c_{kl}\boldsymbol Q_l(i)\boldsymbol Q_l(i-1)\ldots\boldsymbol Q_l(1)\boldsymbol\omega_l(0),\label{wk7}
\end{equation}
where
\begin{equation}
\boldsymbol Q_l(i)=\boldsymbol\Lambda_l^{2i}\boldsymbol{\omega}_l(0)
\big(\boldsymbol{\omega}_l^H(0)\boldsymbol\Lambda_l^{4i-2}\boldsymbol{\omega}_l(0)\big)^{-1}\boldsymbol\omega_l^H(0)\boldsymbol\Lambda_l^{2i-2}.\label{wk6}
\end{equation}
At this point, we need to establish that the norm of $\boldsymbol
S_{D_k}(i)$, for all $i$, is both lower and upper bounded, i.e.,
$0<\parallel\boldsymbol S_{D_k}(i)\parallel<\infty$, for all $i$,
and that $\boldsymbol{\omega}_k(i)=\boldsymbol S_{D_k}(i)\sum_{l\in
\mathcal{N}_k}c_{kl}\boldsymbol{\bar{\omega}}_l(i)$ exponentially
approaches $\boldsymbol{\omega}_{k, \textrm{opt}}(i)$ as $i$
increases. Due to the linear mapping, the boundedness of
$\boldsymbol S_{D_k}(i)$ is equivalent to the boundedness of
$\boldsymbol{\omega}_k(i)$. Therefore, we have upon convergence when
$i \rightarrow \infty$ that
\begin{equation}
\boldsymbol{\omega}_k^H(i)\boldsymbol{\omega}_k(i-1)=\boldsymbol{\omega}_k^H(i-1)\boldsymbol{\omega}_k(i-1).
\end{equation}

Because $\parallel\boldsymbol{\omega}_k^H(i)
\boldsymbol{\omega}_k(i-1)\parallel\leq\parallel\boldsymbol{\omega}_k^H(i-1)\parallel\parallel\boldsymbol{\omega}_k(i)\parallel$
and
$\parallel\boldsymbol{\omega}_k^H(i-1)\boldsymbol{\omega}_k(i-1)\parallel=\parallel\boldsymbol{\omega}_k(i-1)\parallel^2$,
the relation
$\boldsymbol{\omega}_k^H(i)\boldsymbol{\omega}_k(i-1)=\boldsymbol{\omega}_k^H(i-1)\boldsymbol{\omega}_k(i-1)$
implies that
$\parallel\boldsymbol{\omega}_k(i)\parallel\geq\parallel\boldsymbol{\omega}_k(i-1)\parallel$,
and hence we have
\begin{equation}
\parallel\boldsymbol{\omega}_k(\infty)\parallel\geq\parallel\boldsymbol{\omega}_k(i)\parallel\geq\parallel\boldsymbol{\omega}_k(0)\parallel.
\end{equation}
To show that the upper bound
$\parallel\boldsymbol{\omega}_k(\infty)\parallel$ is finite, let us
express the $M \times M$ matrix $\boldsymbol Q_k(i)$ as a function
of the $M \times 1$ vector
$\boldsymbol{\omega}_l(i)=\left[{\begin{array}{c}\boldsymbol{\omega}_{l,1}(i)\\
\boldsymbol{\omega}_{l,2}(i)\\ \end{array}}\right]$ and the $M
\times M$ matrix
$\boldsymbol\Lambda=\left[{\begin{array}{cc}\boldsymbol\Lambda_{l,1}&\\
&\boldsymbol\Lambda_{l,2}\\ \end{array}}\right]$. Substituting the
previous expressions of $\boldsymbol{\omega}_l(i)$ and
$\boldsymbol\Lambda_kl$ into $\boldsymbol Q_l(i)$ as given in
(\ref{wk6}), we obtain
\begin{align}
\boldsymbol Q_l(i)&=\left[{\begin{array}{c}\boldsymbol\Lambda_{l,1}^{2i}
\boldsymbol{\omega}_{l,1}(0)\\ \boldsymbol\Lambda_{l,2}^{2i}\boldsymbol{\omega}_{l,2}(0)\\
\end{array}}\right]\big(\boldsymbol{\omega}_{l,1}^H(0)\boldsymbol\Lambda_{l,1}^{4i-2}\boldsymbol{\omega}_{l,1}(0)\notag\\
&+\boldsymbol{\omega}_{l,2}^H(0)\boldsymbol\Lambda_{l,2}^{4i-2}
\boldsymbol{\omega}_{l,2}(0)\big)^{-1}\left[{\begin{array}{c}
\boldsymbol{\omega}_{l,1}^H(0)\boldsymbol\Lambda_{l,1}^{2i-2}\\
\boldsymbol{\omega}_{l,2}^H(0)\boldsymbol\Lambda_{l,2}^{2i-2}\\
\end{array}}\right]\label{wk9}
\end{align}
Using the matrix identity $(\boldsymbol A +\boldsymbol B)^{-1} =
\boldsymbol A^{-1}-\boldsymbol A^{-1} \boldsymbol B(\boldsymbol I +
\boldsymbol A^{-1}\boldsymbol B)^{-1}\boldsymbol A^{-1}$ in the
decomposed $\boldsymbol Q_l(i)$ in (\ref{wk9}) and making $i$ large,
we get
\begin{equation}
\boldsymbol Q_l(i)=\textrm{diag}(\underbrace{1\ldots1}_D\underbrace{0\ldots0}_{M-D})+O_l\big(\epsilon_l(i)\big),\label{wk8}
\end{equation}
where $\epsilon_l(i)=(\sigma_{r+1}/\sigma_r)^{2i}$, in which
$\sigma_{r+1}$ and $\sigma_r$ are the $(r + 1)$th and the $r$th
largest singular values of $\boldsymbol R_l^{-1/2}(i)\boldsymbol
p_l(i)$, respectively, and $O(\cdot)$ denotes the order of the
argument. Based on (\ref{wk8}), it follows that, for some positive
constant $g$,  we have
$\parallel\boldsymbol{\omega}_l(i)\parallel\leq 1+g\epsilon_l(i)$.
Based on (\ref{wk7}), we obtain
\begin{align}
\parallel\boldsymbol{\omega}_k(\infty)\parallel&\leq\sum_{l\in\mathcal{N}_k}c_{kl}(\parallel\boldsymbol Q_l(\infty)\parallel\ldots\parallel\boldsymbol Q_l(1)\parallel\parallel\boldsymbol Q_l(0)\parallel)\notag\\
&\leq\sum_{l\in\mathcal{N}_k}c_{kl}\bigg(\parallel\boldsymbol{\omega}_l(0)\parallel\prod_{i=0}^{\infty}\big(1+g\epsilon_l(i)\big)\bigg)\notag\\
&=\sum_{l\in\mathcal{N}_k}c_{kl}\bigg(\parallel\boldsymbol{\omega}_l(0)\parallel\exp\bigg(\sum_{i=1}^{\infty}\log\big(1+g\epsilon_l(i)\big)\bigg)\bigg)\notag\\
&\leq\sum_{l\in\mathcal{N}_k}c_{kl}\bigg(\parallel\boldsymbol{\omega}_l(0)\parallel\exp\bigg(\sum_{i=1}^{\infty}g\epsilon_l(i)\bigg)\bigg)\notag\\
&=\parallel\sum_{l\in\mathcal{N}_k}c_{kl}\bigg(\parallel\boldsymbol{\omega}_l(0)\parallel\exp\bigg(\frac{g}{1-(\sigma_{r+1}/\sigma_r)^{2}}\bigg)\bigg).
\end{align}
With the previous development, the norm of
$\boldsymbol{\omega}_k(i)$ is proven to be both lower and upper
bounded. Once this case has been established, the expression in
(\ref{wk3}) converges for a sufficiently large $i$ to the low-rank
Wiener filter. This condition is verified by equating the terms of
(\ref{wk2}), which yields
\begin{equation}
\boldsymbol{\omega}_k(i)=\sum_{l\in\mathcal{N}_k}c_{kl}\bigg(\boldsymbol
R_l^{-1/2}(i)\boldsymbol\Phi_{l,1}\boldsymbol
\Lambda_{l,1}\boldsymbol\Phi_{l,1}^H\boldsymbol
p_l(i)+O_l\big(\epsilon_l(i)\big)\bigg)
\end{equation}
where $\boldsymbol\Phi_{l,1}$ is a $M \times D$ matrix with the $D$ largest
eigenvectors of $\boldsymbol R_l(i)$, and $\boldsymbol\Lambda_{l,1}$ is a $D
\times D$ matrix with the largest eigenvalues of $\boldsymbol R_l(i)$.

\section{Simulation results}

In this section, we investigate the performance of the proposed DRJIO--NLMS and
DRJIO--RLS algorithms for distributed estimation in two scenarios: wireless
sensor networks and smart grids.

\subsection{Wireless Sensor Networks}

In this subsection, we compare the proposed DRJIO--NLMS and
DRJIO--RLS algorithms with the distributed NLMS algorithm
(normalized version of \cite{Lopes2}), distributed RLS algorithm
\cite{Cattivelli2}, Krylov subspace NLMS \cite{Chouvardas1} and
distributed principal subspace estimation \cite{Li3}, based on their
MSE performance.

\begin{figure}
\begin{center}
\def\epsfsize#1#2{0.9\columnwidth}
\epsfbox{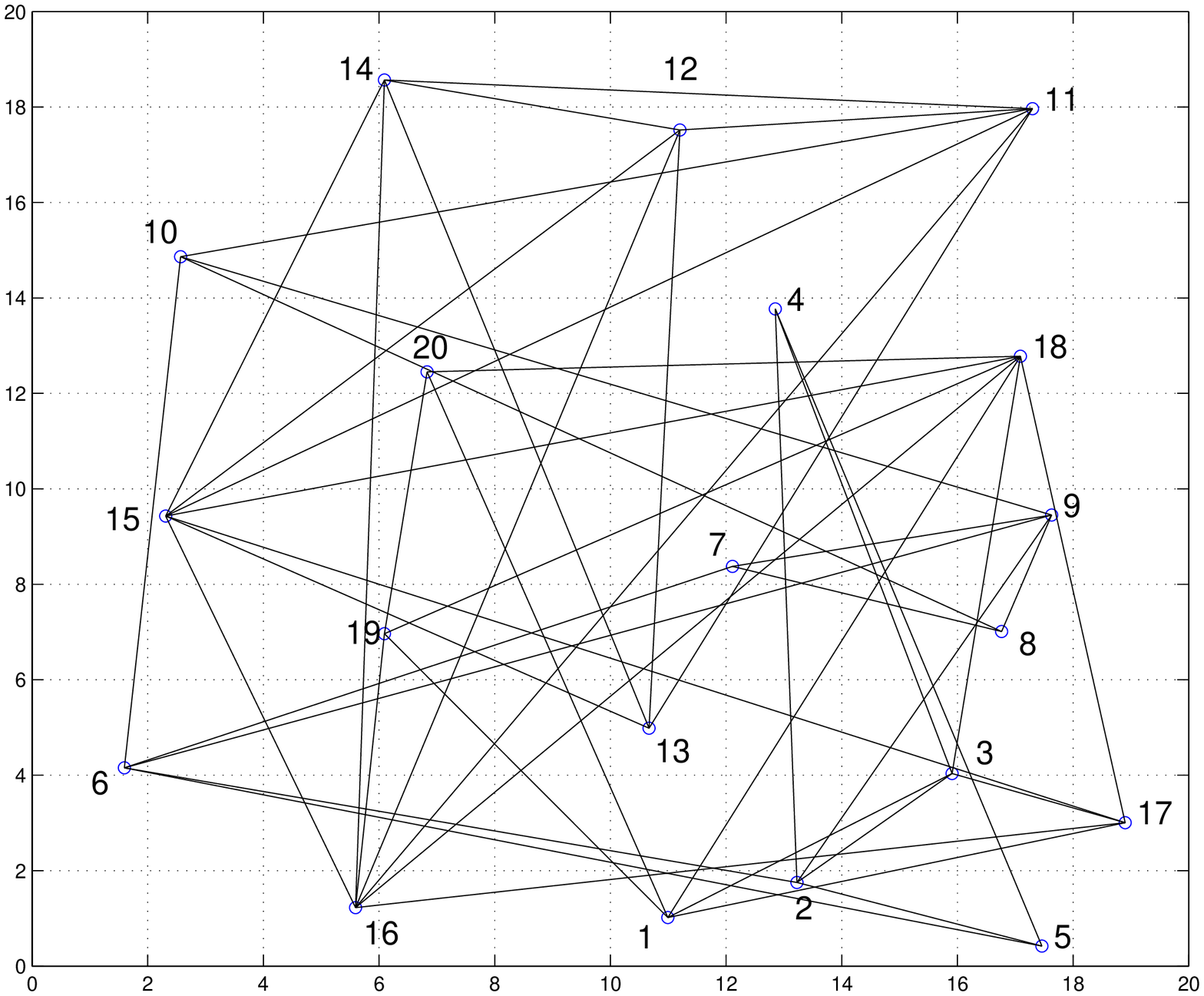}
\caption{\footnotesize Network topology with $N=20$ nodes}
\label{fig_net}
\end{center}
\end{figure}

With the network topology structure outlined in Fig. \ref{fig_net} with $N=20$
nodes, we consider numerical simulations under three scenarios for the
parameter vector ${\omega}_o$:
\begin{itemize}
\item Full--rank system with $M$=20
\item Sparse system with $M$=20 ($D$ non-zero coefficients and $M-D$ zeros coefficients)
\item Full--rank system with $M$=60
\end{itemize}
The input signal is generated as ${\boldsymbol x_k(i)}=[x_k(i)\ \ \
x_k(i-1)\ \ \ ...\ \ \ x_k(i-M+1)]$  and
$x_k(i)=u_k(i)+\alpha_kx_k(i-1)$, where $\alpha_k$ is a correlation
coefficient and $u_k(i)$ is a white noise process with variance
$\sigma^2_{u,k}= 1-|\alpha_k|^2$, to ensure the variance of
${\boldsymbol x_k(i)}$ is $\sigma^2_{x,k}= 1$.  {In particular, this
application requires the estimation of a set of parameters that
could be modeled as a finite-impulse response (FIR) filter (related
to a moving average (MA) model). Furthermore, the algorithms would
also work with an input generated by an auto-regressive (AR) model
but their performance would depend on the condition number and the
rank of the correlation matrix of the input data.} The noise samples
are modeled as complex Gaussian noise with variance $\sigma^2_{n,k}=
0.001$. We have adopted the regularization parameters $\gamma =
0.02$ and $\delta = 0.01$ in all examples. We have also evaluated
the impact of different values of regularization parameters and the
results indicate that the performance of the algorithms degrades
when the parameters are not well chosen. Moreover, the optimized
values work very well for a wide range of scenarios and values of
noise variance. We assume that the network has error--free
transmission between linked nodes.

\begin{figure}
\begin{center}
\def\epsfsize#1#2{1.0\columnwidth}
\epsfbox{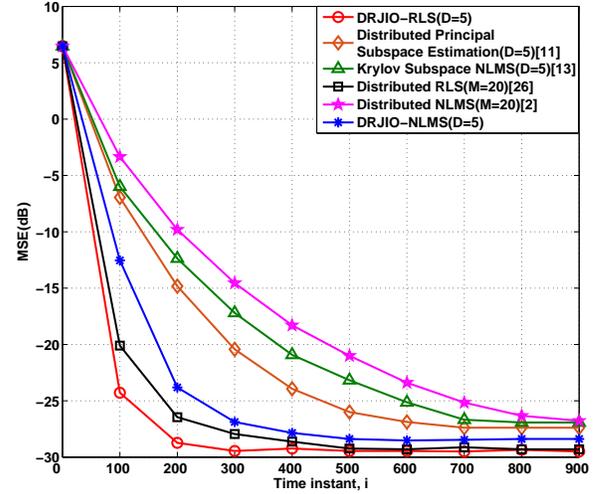}\caption{\footnotesize Full--rank system
with $M$=20}  \label{fig6:fig3}
\end{center}
\end{figure}

The step size $\mu_0$ for the distributed NLMS algorithm, Krylov
subspace NLMS, distributed principal subspace estimation and
DRJIO--NLMS is set to 0.15 and $\eta_0$ is set to $0.5$. For the
distributed RLS algorithm and DRJIO--RLS algorithm, the forgetting
factor $\lambda$ is equal to 0.99 and $\delta$ is set to 0.11. In
Fig. \ref{fig6:fig3}, we compare the proposed DRJIO--NLMS and
DRJIO--RLS algorithms with the existing strategies using the
full--rank system with $M$=20 and $D$=5. The dimensionality
reduction matrix $\boldsymbol S_{D_k}(0)$ is initialized as
$[\boldsymbol I_D \ \ \boldsymbol 0_{D,M-D}]^T$.

We observe that the proposed DRJIO--RLS algorithm has the best
performance when compared with other algorithms, while the proposed
DRJIO--NLMS algorithm also has a better performance, which is very
close to the distributed RLS algorithm. {The superior performance of
DRJIO-RLS can be explained by the fact that the convergence rate or
learning speed of adaptive algorithms depends on the number of
parameters that need to be estimated. This is well known in adaptive
signal processing \cite{Haykin}. For instance, when we compare the
DRJIO-RLS and the full-rank RLS, the difference is that the proposed
DRJIO-RLS estimates the unknown parameters using a reduced dimension
and retaining the most relevant features of the data. As a result,
the DRJIO-RLS converges faster than the standard RLS algorithm.
However, its complexity is an order of magnitude lower than those of
the distributed RLS algorithm and the DRJIO--RLS algorithm.}

\begin{figure}
\begin{center}
\def\epsfsize#1#2{1.0\columnwidth}
\epsfbox{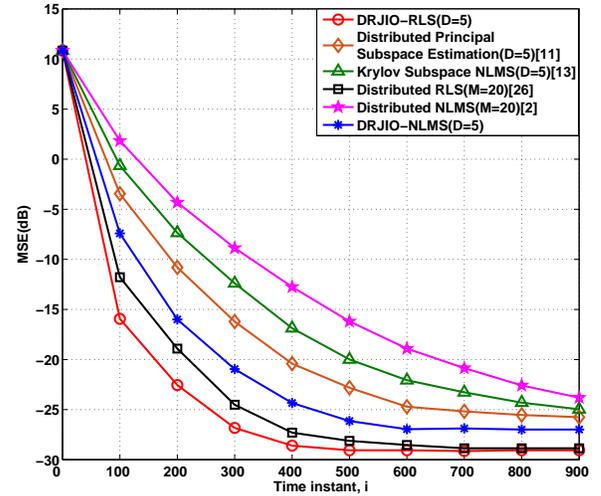} \caption{\footnotesize
Full--rank system with $M$=60}  \label{fig6:fig5}
\end{center}
\end{figure}

When the full--rank system $M$ increases to $60$, Fig.
\ref{fig6:fig5} illustrates that, the proposed DRJIO--RLS algorithm
still has the best performance, while DRJIO--NLMS algorithm also
shows a fast convergence rate, which is comparable to the
distributed RLS algorithm. For the distributed NLMS, Krylov subspace
NLMS and distributed principal subspace estimation algorithms, their
convergence speed is much lower.

\begin{figure}[!htp]
\begin{center}
\def\epsfsize#1#2{1.0\columnwidth}
\epsfbox{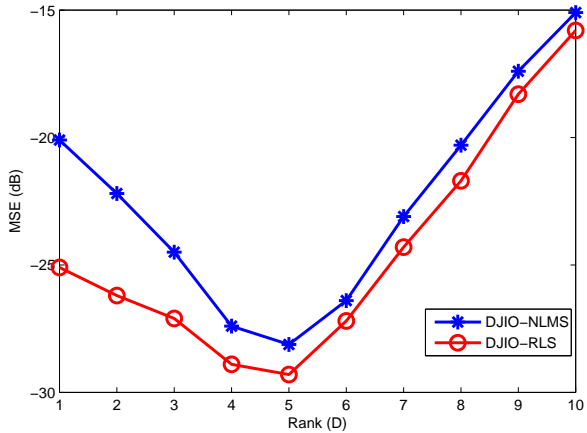} \caption{\footnotesize MSE performance versus
rank $D$ for a sparse system with $M$=100}\vspace{-1em}
\label{fignew:fig7}
\end{center}
\end{figure}

\begin{figure}[htb!]
\begin{center}
\def\epsfsize#1#2{1.0\columnwidth}
\epsfbox{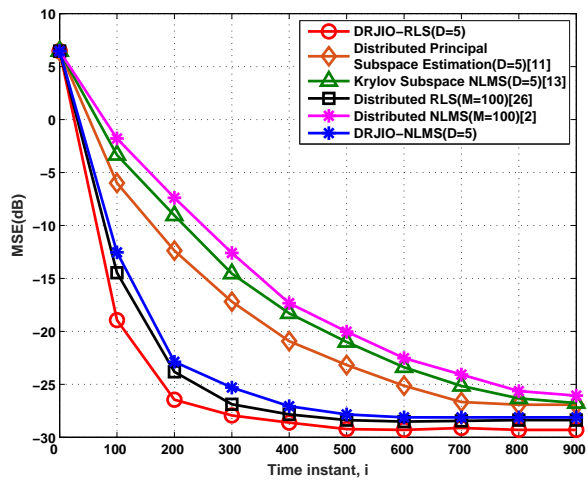} \caption{\footnotesize Sparse system with
$M$=100} \label{fig6:fig4}
\end{center}
\end{figure}

In a sparse system scenario with $M=100$, we first evaluate the MSE
performance versus the rank $D$ and then we assess the MSE
performance versus the number of iterations, as shown in Figs.
\ref{fignew:fig7} and \ref{fig6:fig4}, respectively. In particular,
the curves illustrating the MSE performance versus the rank $D$ are
obtained after $500$ iterations for a range of $D$ between $1$ and
$10$. The results depicted in Fig. \ref{fignew:fig7} indicate that
the best rank for both DJRIO--NLMS and DJRIO--RLS algorithms
corresponds to $D=5$ and that the MSE performance gradually degrades
for other values. The rank $D$ should be carefully selected as it
affects the performance of these algorithms and determines the
number of parameters that should be exchanged between nodes.
Moreover, we have considered $D=5$ for assessing the MSE performance
versus the number of iterations for the proposed and other existing
algorithms, as shown in Fig. \ref{fig6:fig4}. The results indicate
that the proposed DRJIO--RLS and DRJIO--NLMS algorithms have a more
pronounced performance advantage over the distributed NLMS, the
Krylov subspace NLMS and the distributed principal subspace
estimation algorithms. Specifically, the proposed DRJIO--NLMS
algorithm performs very close to the distributed RLS algorithm and
outperforms the other analyzed algorithms.

\begin{figure}[p]
\begin{center}
\def\epsfsize#1#2{1.0\columnwidth}
\epsfbox{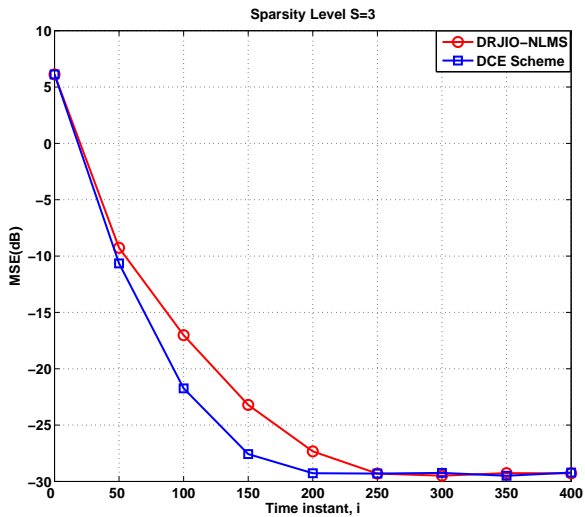} \caption{\footnotesize DRJIO--NLMS vs DCE
scheme with sparsity level S=3}\vspace{-1em}  \label{fig6:fig6}
\end{center}
\end{figure}

\begin{figure}[!htp]
\begin{center}
\def\epsfsize#1#2{1.0\columnwidth}
\epsfbox{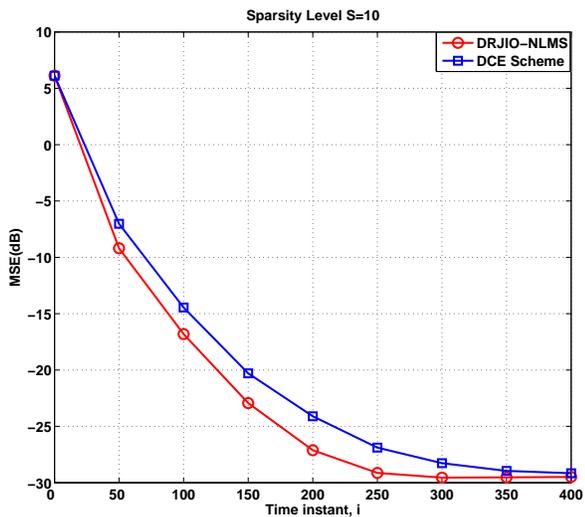} \caption{\footnotesize DRJIO--NLMS vs DCE
scheme with sparsity level S=10}\vspace{-1em} \label{fig6:fig7}
\end{center}
\end{figure}

In the last example on wireless sensor networks, we compare the
performance between the proposed DRJIO--NLMS and the DCE scheme in
\cite{dce}, under different sparsity level scenarios. The step size
for both algorithms is set to 0.3 and the $\eta_0$ for DRJIO--NLMS
is set to 0.5. The length of the unknown parameter $\boldsymbol
\omega_0$ is 20 and $D=10$. For the first scenario, the number of
non--zero coefficients in the unknown parameter is 3 and for the
second scenario, the number of non--zero coefficients is set to 10.
The comparison results are shown in Fig. \ref{fig6:fig6} and
\ref{fig6:fig7}. It is clear that in a very sparse system, the
proposed DCE scheme outperforms the DRJIO--NLMS algorithm. With the
decrease of the system sparsity level, the proposed DRJIO--NLMS
algorithm outperforms the DCE scheme. The results of \ref{fig6:fig6}
and \ref{fig6:fig7} indicate that the proposed DRJIO--NLMS algorithm
is superior to the DCE scheme when the level of sparsity is not very
high. The computational complexity of DRJIO-NLMS is much lower than
the DCE scheme because the latter requires a basis pursuit algorithm
to reconstruct the full-dimension estimator and DRJIO-NLMS employs a
rank-$D$ approximation based on a simple matrix-vector
multiplication.

\subsection{Smart Grids}

In order to test the proposed algorithms in a possible smart grid scenario, we
consider the Hierarchical IEEE 14--bus system which has been proposed in
\cite{Ma}, where 14 is the number of substations. At every time instant $i$,
each bus $k,k=1,2, \ldots, 14 ,$ takes a scalar measurement $d_k(i)$ according
to
\begin{equation}
{d_k(i)}= {X_k \big({\boldsymbol \omega}_0(i)\big)+ n_k(i)},~~~
k=1,2, \ldots, 14 \label{Eqn6:z_k},
\end{equation}
where $\boldsymbol \omega_0(i)$ is the state vector of the entire
interconnected system, $X_k({\boldsymbol \omega}_0(i))$ is a
nonlinear measurement function of bus $k$. The quantity ${n_k(i)}$
is the measurement error with mean equal to zero and which
corresponds to bus $k$.

\begin{figure}
\begin{center}
\def\epsfsize#1#2{0.8\columnwidth}
\epsfbox{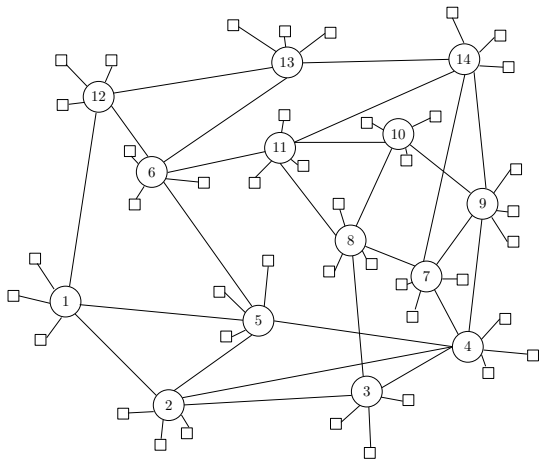}\caption{\footnotesize
Hierarchical IEEE 14--bus system}
\label{fig6:sg}
\end{center}
\end{figure}

We focus on the linearized DC state estimation problem. We assume that each bus
connects and measures the state of three users. As a result, for the IEEE--14
bus system, there will be 42 users in the system. The system is built with 1.0
per unit (p.u) voltage magnitudes at all users and j1.0 p.u. branch impedance.
Then, the state vector ${\boldsymbol \omega}_0(i)$ is taken as the voltage
phase angle vector ${\boldsymbol \omega}_0$ for all users. Initially, each bus
only knows the voltage phase angle of the three users connected to it. With the
help of distributed estimation algorithms, each bus is supposed to estimate the
state of the voltage phase angles for all users in the system. Therefore, the
nonlinear measurement model for state estimation (\ref{Eqn6:z_k}) is
approximated by
\begin{equation}
{d_k(i)}= {\boldsymbol x_k^H(i)\boldsymbol \omega_0+ n_k(i)},~~~
k=1,2, \ldots, 14 ,
\end{equation}
where ${\boldsymbol x}_k(i)$ is the measurement Jacobian vector for
bus $k$. Then, the aim of the distributed estimation algorithm is to
compute an estimate of ${\boldsymbol \omega}_0$, which can minimize
the cost function given by
\begin{equation}
{J_{\boldsymbol\omega_k(i)}({\boldsymbol \omega_k(i)})} =
{\mathbb{E} |{ d_k(i)}-{\boldsymbol x_k^H(i)}{\boldsymbol\omega_k(i)}}|^2.
\end{equation}
and the global network cost function is described by
\begin{equation}
{J_{\omega}({\boldsymbol \omega})} = \sum_{k=1}^{N}{\mathbb{E} |{ d_k(i)}-
{\boldsymbol x_k^H(i)}{\boldsymbol \omega}|^2}.\label{Eqn6:cost function2}
\end{equation}
We compare the proposed algorithms with the
$\mathcal{M}$--$\mathcal{CSE}$ algorithm \cite{Xie}, the distributed RLS algorithm
\cite{Cattivelli2}, the distributed NLMS algorithm (normalized version of \cite{Lopes2}) and
distributed principal subspace estimation \cite{Li3} in terms of MSE performance.
The MSE comparison is used to determine the accuracy of the algorithms and the rate of convergence. We define the Hierarchical IEEE--14 bus system as in Fig.
\ref{fig6:sg}.

\begin{figure}
\begin{center}
\def\epsfsize#1#2{1.0\columnwidth}
\epsfbox{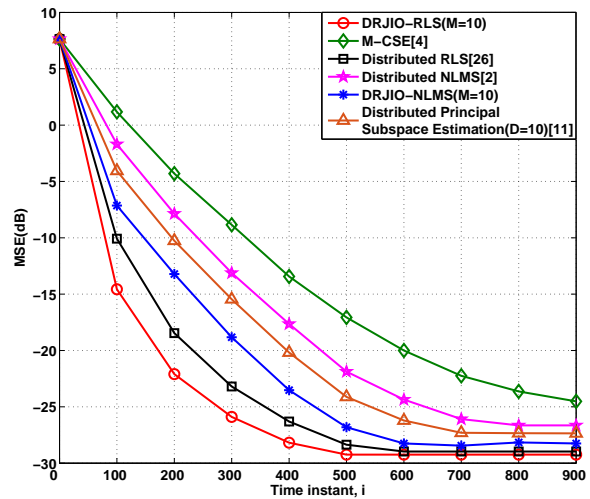} \caption{\footnotesize
MSE  performance for smart grids} \label{fig6:sg2}
\end{center}
\end{figure}

All buses are corrupted by additive white Gaussian noise with variance $\sigma^2_{n,k}=0.001$. The step size for the distributed NLMS \cite{Lopes2} and the proposed DRJIO--NLMS algorithms is
$\mu=0.15$ and $\eta_0$ is set to 0.5. The parameter vector $\boldsymbol \omega_0$ is set to an
all--one vector with size $42\times 1$. For the distributed RLS, DRJIO--RLS
algorithms the forgetting factor $\lambda$ is set to 0.99 and
$\delta$ is equal to 0.11. The reduced dimension $D$ is set to 10 for both DRJIO--RLS and DRJIO--NLMS algorithm. The results are averaged over 100 independent runs. We simulate the proposed algorithms for smart grids under a static scenario.

From Fig. \ref{fig6:sg2}, it can be seen that the proposed DRJIO--RLS algorithm has the best
performance, and significantly outperforms the distributed NLMS \cite{Lopes2} and the $\mathcal{M}$--$\mathcal{CSE}$ \cite{Xie}
algorithms. The DRJIO--NLMS is slightly worse than distributed RLS algorithm
\cite{Cattivelli2}, but better than the distributed NLMS and $\mathcal{M}$--$\mathcal{CSE}$ algorithms. In addition, the proposed DRJIO--NLMS and DRJIO--RLS algorithms can compress the data to be transmitted from each node from $M$ to $D$, resulting in reduced bandwidth requirements. These algorithms are also important tools for dealing with large sets of data which exhibit some form of redundancy, sparsity and are compressible.

\section{Conclusions}
In this paper, we have proposed a novel distributed low-rank scheme
along with efficient algorithms for distributed estimation in
wireless sensor networks and smart grids. Simulation results have
shown that the proposed DRJIO--RLS has the best performance, while
DRJIO--NLMS algorithm has a better performance and lower cost than
existing algorithms in all the three scenarios considered. We have
also compared the proposed algorithms with the DCE scheme, which was
presented in \cite{dce}, for systems with different levels of
sparsity. Furthermore, the proposed scheme requires the transmission
of only $D$ parameters instead of $M$, resulting in higher bandwidth
efficiency than standard schemes.

{\begin{appendix}[Reconstruction using a rank-$D$ approximation]

In this appendix, we show how the reconstruction of the
full-dimension estimator ${\boldsymbol \omega_k}(i)$ can be carried
out using a rank-$D$ approximation with the low-rank estimator
$\bar{\boldsymbol \omega_k}(i)$, i. e.,
\begin{equation}
{\boldsymbol \omega_k}^{(D)}(i) = {\boldsymbol S}_{D_k}(i)
\bar{\boldsymbol \omega}_k(i), \label{desired_result}
\end{equation}
In order to show the above relation, we consider the expression of
the low-rank estimator given by
\begin{equation}
\begin{split}
\bar{\boldsymbol \omega_k}(i) & = \bar{\boldsymbol R}_k^{-1}(i) \bar{\boldsymbol p}_k(i) \\
& = \Big({\boldsymbol S}_{D_k}^H(i) {\boldsymbol R}_k(i) {\boldsymbol S}_{D_k}(i)
\Big)^{-1} {\boldsymbol S}_{D_k}^H(i) {\boldsymbol p}_k(i).
\end{split}
\end{equation}
and the Wiener filter given by
\begin{equation}
\begin{split}
{\boldsymbol \omega_k}(i) & = {\boldsymbol R}_k^{-1}(i) {\boldsymbol p}_k(i).
\end{split}
\end{equation}
Using the fact that the low-rank estimator converges to the low-rank
Wiener filter, ${\boldsymbol S}_{D_k}(i)$ converges to a $M \times
D$ matrix with the eigenvectors ${\boldsymbol \Phi}_D$ and an
eigenvalue decomposition of ${\boldsymbol R}_k(i)= {\boldsymbol
\Phi}_N {\boldsymbol \Lambda}_N {\boldsymbol \Phi}_N^H =\sum_{n=1}^N
\lambda_n(i) {\boldsymbol \phi}_n(i) {\boldsymbol \phi}_n^H(i)$,
where  ${\boldsymbol \Lambda}_N$ and ${\boldsymbol \Phi}_N$ are the
$N \times N$ diagonal matrix with the eigenvalues and the  $N \times
N$ unitary matrix with the eigenvectors of ${\boldsymbol R}_k(i)$,
respectively, $\lambda_n(i)$ is the $n$th eigenvalue and
${\boldsymbol \phi}_n(i)$ is the $n$th eigenvector of ${\boldsymbol
R}_k(i)$, we have
\begin{equation}
\begin{split}
\bar{\boldsymbol \omega_k}(i) & = \Big({\boldsymbol S}_{D_k}^H(i) \sum_{n=1}^{N}
\lambda_n(i) {\boldsymbol \phi}_n(i) {\boldsymbol \phi}_n^H(i)
{\boldsymbol S}_{D_k}(i) \Big)^{-1} {\boldsymbol S}_D^H(i) {\boldsymbol p}_k(i)\\
& = ({\boldsymbol \Phi}_{D}^H {\boldsymbol \Phi}_{N} \Lambda_N
{\boldsymbol \Phi}_{N}^H {\boldsymbol \Phi}_{D})^{-1} {\boldsymbol \Phi}_D^H {\boldsymbol p}_k(i)\\
& = \Lambda_D^{-1} {\boldsymbol \Phi}_D^H {\boldsymbol p}_k(i),
\label{eig_rr}
\end{split}
\end{equation}
then, multiplying ${\boldsymbol S}_{D_k}(i)={\boldsymbol \Phi}_D$ on
both sides, we obtain
\begin{equation}
\begin{split}
{\boldsymbol S}_{D_k}(i)\bar{\boldsymbol \omega_k}(i)&  =
{\boldsymbol S}_{D_k}(i) \Lambda_D^{-1} {\boldsymbol \Phi}_D^H {\boldsymbol p}_k(i)\\
&  = {\boldsymbol \Phi}_D \Lambda_D^{-1} {\boldsymbol \Phi}_D^H {\boldsymbol p}_k(i)\\
& ={\boldsymbol \omega_k}^{(D)}(i),
\end{split}
\end{equation}
where ${\boldsymbol R}_k^{(D)}(i) = {\boldsymbol \Phi}_D \Lambda_D
{\boldsymbol \Phi}_D^H$ is a rank-$D$ approximation of ${\boldsymbol
R}_k(i)$ and ${\boldsymbol \omega_k}^{(D)}(i) = {\boldsymbol \Phi}_D
\Lambda_D^{-1} {\boldsymbol \Phi}_D^H {\boldsymbol p}_k(i)$ is the
rank-$D$ approximation of ${\boldsymbol \omega_k}(i)$, which gives
us the relation in (\ref{desired_result}). Note that when $D = M$,
the $D$-rank approximation yields the full-rank Wiener filter.

\end{appendix}}

\bibliographystyle{IEEEtran}
\bibliography{reference}

\end{document}